\documentclass[english,a4paper]{article}
\usepackage{amsmath}
\usepackage{graphicx}
\usepackage{amssymb}
\usepackage{amsfonts}

\newtheorem{theorem}{Theorem}

\newtheorem{coro}{Corollary}
\newtheorem{prin}{Principle}

\newtheorem{lemma}{Lemma}

\newtheorem{obs}{Observation}

\usepackage{babel}
\begin{document}

\title{The form of the rate constant for  elementary reactions at equilibrium  from MD: framework and proposals for thermokinetics}

\author{Christopher G. Jesudason
\thanks{on leave from Chemistry Dept., University of Malaya, 50603 Kuala Lumpur, Malaysia. Fax:603-79674193, Email:jesu@um.edu.my}\\
 {\normalsize Dept. of Chemistry,  } \\
 {\normalsize University of Maine, Orono}\\
 {\normalsize Maine ME04469,\,\,U.S.A.}}
\date{\normalsize July, 2006}

\vspace{.5cm}
\maketitle
\begin{abstract}
The rates or formation  and concentration distributions of a dimer reaction showing hysteresis behavior are examined in an ab initio chemical reaction designed as elementary and where the hysteresis structure precludes the formation of transition states (TS) with pre-equilibrium and internal sub-reactions. It was discovered that  the  the reactivity coefficients, defined as a measure of departure from the zero density rate constant for the forward and backward steps had a ratio that was equal to   the activity coefficient ratio for the product and reactant species. This surprising result, never formally noticed nor incorporated in elementary rate expressions  over approximately one and a half centuries of quantitative chemical kinetics measurement and calculation  is accepted axiomatically and leads to an outline of a theory for  the form of the rate constant, in any one given substrate - here the vacuum state. A  major deduction  is that the long-standing definition of the rate constant used for over a century and a half  for elementary reactions is not complete, where  previous works almost always implicitly refer to the zero density limit for strictly irreducible elementary reactions without any any attending concatenation of side-reactions. This is shown directly from MD simulation,where for specially designed elementary reactions without any transition states, density dependence of reactants and products always feature, in contrast to current practice.    It is argued that the  rate constant expression  without reactant and product  dependence is due to  historical conventions  for strictly elementary reactions.  From the above  observations, a theory is developed with the aid of some proven   elementary theorems in thermodynamics, and 
expressions under different state conditions are derived whereby a feasible    experimental and computational method  for determining  the activity coefficients
from the rate constants may be obtained under various approximations and conditions. Elementary relations for subspecies equilibria and its relation to the bulk activity coefficient are discussed. From one choice of reaction conditions, estimates of activity coefficients are given which are in at least semi-quantitative agreement with the data for non-reacting Lennard-Jones (LJ) particles for the atomic component. The theory developed is applied to ionic reactions  where the standard Br\"onsted-Bjerrum rate equation and exceptions to this are rationalized, and by viewing ion association as a n-meric process,  ion activity coefficients  may in principle be determined under varying applicable assumptions. 
\end{abstract}
\section{Introduction}
\label{sec:1} Previous work has detailed a hysteresis model of a simple dimer reaction \cite{cgj7,jescf1,cgj6} where the coordinates for  molecular formation $r_f$ and breakdown $r_b$  are not at the  same vicinity, as shown in Fig.(\ref{fig:1}). 
If both these points coincided, then one could in principle define a volume region about the point $r_f=r_b$  that would serve as the transition state pre-equilibrium $TS$, suggesting a composite reaction such as 
\begin{equation}\label{add1}
	2\text{A}\rightleftharpoons TS \rightarrow \text{A}_2 
\end{equation}
which is therefore not strictly elementary since (\ref{add1}) is a summation of elementary steps to yield a net reaction. The current model precludes such a possibility , implying a strictly elementary reaction process. Previous models, especially that of Eyring postulated the pre-equilibrium TS which was also interpreted experimentally \cite{pol1} but the most recent developments in some cases by-passes this development \cite{mil1,pnas1}.  In the theory that follows, the $TS$ refers to the point in state space whose neighborhood always includes product and reactant states.   The method for treating changes of potential via switch mechanisms were also outlined\cite{cgj7,jescf1,cgj6}, together  with an algorithm which was implemented to   conserve energy and momentum at high energy regions with steep potential gradients.  This work  focuses  primarily on
the form of the elementary rate constants for the reaction, and presents a way of determining   activity and {\itshape reactivity} coefficients; the two concepts are different, and will be defined in this work, where under certain conditions, the activity coefficients might be determined. These two types of coefficients are discriminated by first introducing an ansatz of the 
 dependency of the
reactivity coefficients on the rate constants and comparing the results with the actual kinetics of the  simulation, where it was discovered that there were variations  of the rate constant with concentration. Section (\ref{sec:3}) and its subsections give details of the methods used to determine the extent of variation. The ratio of the reactivity coefficient $\Phi_k$ agreed numerically  with the results from the activity factor $\Phi_e$, which is determined  independently  by extrapolation of   the variation of the concentration equilibrium
ratio determined directly from  simulation. This observation is used to develop a theory for the form   elementary  reaction rate constants  take,  which does not accord with standard assumptions. The above 
objectives are realized by first describing the model of the elementary reaction in Section (\ref{sec:2}). The empirical result of the equivalence of the reactivity and activity coefficient ratios is developed in Section  (\ref{sec:3}) and the subsections where the consequences are discussed in depth. The uniqueness of the activity coefficient and various other associated results appear in Section(\ref{sub:3.2}).  Verification of the theories concerning the role of activity coefficients in elementary reactions is given in reference to results from the literature for non-reacting LJ particles, where it is argued that the residual Helmholtz energy $A_{res}$ is a better measure for computing the activity coefficient for multi-component system than the residual free energy $G_{res}$ (Sections (\ref{subsub:3.3.1}-\ref{sub:3.6.3})). In either case, the activity coefficient  for the unreacted particle  increases with system concentration and is positive. Of two possibilities for the estimate of the activity coefficient derived from energy considerations of the dimer and single  particle trajectory along the reaction coordinate, only one accords with observation derived from simulation data from the literature for the activity coefficient of the atom, with semi-quantitative agreement. From the form of the activity/availability coefficients, rudimentary mechanisms involving single and double stages (Sections (\ref{sub:3.6.1}-\ref{sub:3.6.3}))are proposed for elementary reactions. It is suggested that the 2-stage mechanisms is the more appropriate, and an interpretation of the Br\"onsted-Bjerrum (BB) rate equation is made on the basis of the two-stage model (Sections (\ref{sub:3.6.4}-\ref{sub:3.6.5})). The conclusion (Section(\ref{sec:4}))brings into focus all the details of the   preceding sections.
\section{The Model}

\label{sec:2}

The simulation model is  a dimeric particle reaction 
\begin{equation}
2\text{A}\rightleftharpoons\text{A}_{2}\label{e1}\end{equation}
  at the Lennard-Jones (LJ)  supercritical regime ($T^\ast=8.0, \,\,0.03<\rho<1.1$) in a range of equilibrium fluid states. Details of the mechanism have already been described and will not be repeated here \cite{cgj8,cgj10}.
\setcounter{figure}{0} %
\begin{figure}[htbp]
\begin{center}\includegraphics[%
  width=11cm]{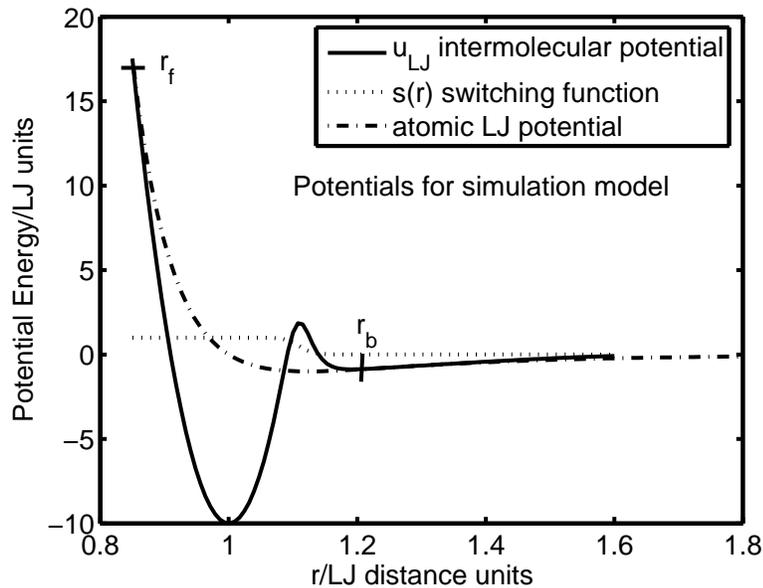} \end{center}
\caption{Potentials used for this work}
\label{fig:1} 
\end{figure}
In the current study, the potentials  given in Fig.~\ref{fig:1}
are used. In this model, 2 free atoms "`react"' at $r_f$ where a switch mechanism converts the potential to the harmonic intermolecular potential. The molecule "`breaks"' at $r_b$ where the potential reverts to the LJ type. At these points, a specialized algorithm was used to conserve energy and momentum and this too has been described in detail \cite{cgj6,cgj8}. The
free atoms A interact with all other particles (whether A or A$_{2}$)
via a Lennard-Jones spline potential and this type of potential has
been described in much detail \cite{Haf7}. An atom at
a distance $r$ to another particle possesses a mutual potential energy
$u_{LJ}$ where \begin{align}\label{e3late}
u_{LJ} & =4\varepsilon\left[\left(\frac{\sigma}{r}\right)^{12}-\left(\frac{\sigma}{r}\right)^{6}\right]\text{
\ \ \ \ \ \ \ \ \ \ \ \ \ \ \ \ \ \ \  for }r\leq r_{s}\\ 
u_{LJ} & =a_{ij}(r-r_{c})^{2}+b_{ij}(r-r_{c})^{3}\text{
\ \ \ \ \ \ \ \ \ \  for }r_{s}\leq r\leq r_{c}\nonumber \\
u_{LJ} & =0\text{ \ \ \ \ \ \ \ \ \ \ \ \ \  for }r>r_{c}\nonumber \end{align}
 and \ where $r_{s}=(26/7)^{\frac{1}{6}}\sigma$ \cite{Haf7}. At $r_f$,
the potential is switched  to the molecular
potential given by

\begin{equation}
u(r)=u_{vib}(r)s(r)+u_{LJ}\left[1-s(r)\right]\label{e4}
\end{equation}
where $u_{vib}(r)$ is the vibrational potential given
by eq.(\ref{eq:5}) below and the switching function $s(r)$ has the
form given by eq.(\ref{eq:6}).
 \begin{equation}
s(r)=\frac{1}{1+\left(\frac{r}{r_{sw}}\right)^{n}}\label{eq:6}
\end{equation}
 where \[
\left\{ \begin{array}{ll}
s(r) & \rightarrow1\text{ \ \ \ \ \  if }r<r_{sw}\\
s(r) & \rightarrow0\text{ \ \ \ \ \  for }r>r_{sw}\end{array}\right..\]
. The switching function becomes effective when the distance between
the atoms approach the value $r_{sw}$ (see Fig.~(\ref{fig:1}).
The intramolecular vibrational potential $u_{vib}(r)$\  for a molecule
is given by \begin{equation}
u_{vib}(r)=u_{0}+\frac{1}{2}k(r-r_{0})^{2}.\label{eq:5}\end{equation}
LJ reduced units are used throughout
this work unless stated otherwise. The details of the parameters have been given elsewhere \cite{cgj8} but they include the following:
 \newline  $u_{0}=-10,r_{0}=1.0,k\sim2446$
(exact value is determined by the other input parameters),\,\,$n=100,r_{f}=0.85,r_{b}=1.20,\mbox{and}\; r_{sw}=1.11$.

\section{Thermodynamic results from equilibrium mixtures}\label{sec:3} 
Details of the runs have been described  elsewhere \cite{cgj8}. Typical runs of 10 million ($10M$) time steps were performed at each general system particle
density $\rho$ (where $\rho$ refers to  the particle which is either free or  part of a molecule),
where the first 200,000 steps were discarded so that proper equilibration
could be achieved for our data samples. The sampling methods have
been previously described \cite{Haf7} where sampling of all data
variables were done each $20^{th}$ time step and where the data were averaged and written  into a dump file of  100 dumps for the 10M  million time steps. The averaged values in each dump were further averaged to yield the final averages and standard errors. Dynamical quantities however had to be
sampled at each time step $\delta t^{*}=0.00005$. In view of the abnormally high temperatures --not hitherto encountered  in almost  all simulation studies-- this time step value  was  found to be {\itshape not} too small. The system
was thermostatted at the ends of the MD cell only, but very similar unpublished 
results with less variable fluctuation  was obtained by thermostatting
each layer with strict conservation of momentum during the thermostatting
process \cite{Tamio9} which involves solving coupled equations for
each thermostatted layer. But it was desired to mimic the actual experimental situation where the reservoirs occurred at the boundary location of the system.It is surmised that reservoir boundary conditions could play a pivotal role in determining the product outcomes of  reactions sensitive to energy fluctuations.  The term $sde$ in the figure captions refer to the standard error.

\subsection{Equilibrium constants}\label{sub:3.1}
\begin{figure}[htbp]

\begin{center}\includegraphics[
  width=10cm]{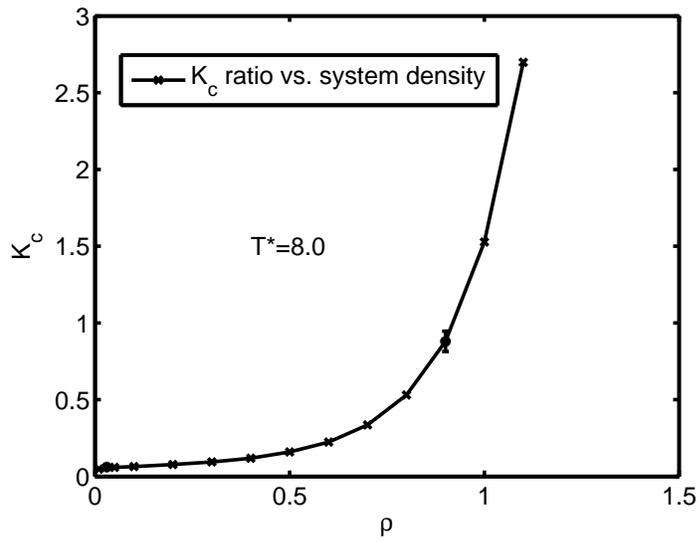} \end{center}

\caption{Variation of concentration ratio $K_{c}$ with $\rho$, the system
number density at LJ temperature $T^{*}=8.0$ with $sde=3$ at $\rho=.03$
and $sde=50$ at $\rho=.9$ }

\label{fig:4} 
\end{figure}

Two independent methods ( (i) and (ii) ) were used to  confirm the thermodynamical results.
\newline (i) {\itshape Time independent distribution sampling method}
\newline 
In order to find the thermodynamic equilibrium constant, $K_{eq}$,
the following procedure was adopted. The concentration ratio, $K_{c}$
defined as
 \begin{equation}
K_{c}=\frac{x_{\text{A}_{2}}}{x_{\text{A}}^{2}}
\end{equation}
 was determined as a function of average system density, $\rho$ \ where
the $x$'s represent number density concentrations of the species indicated by subscript for the reaction (\ref{e1}). At very small
densities, the system becomes an 'ideal' mixture, but as illustrated
previously \cite{cgj7}, the limit of the potentials cannot be the same as the
isolated potentials used in the MD calculations, since if this were
the case, all the molecules would break up, yielding a net zero value
for the equilibrium constant at the limit of zero density. As another
project, it would be of interest to determine the limit at which the
equilibrium regime breaks down in this thermostatted system, but there
may well be technical difficulties involved in computations of very
low density systems. The plot of $K_{c}=K_{c}(\rho)$\ is shown in
Fig.~(\ref{fig:4}).

\begin{figure}[htbp]
\begin{center}
\includegraphics[ width=11cm]{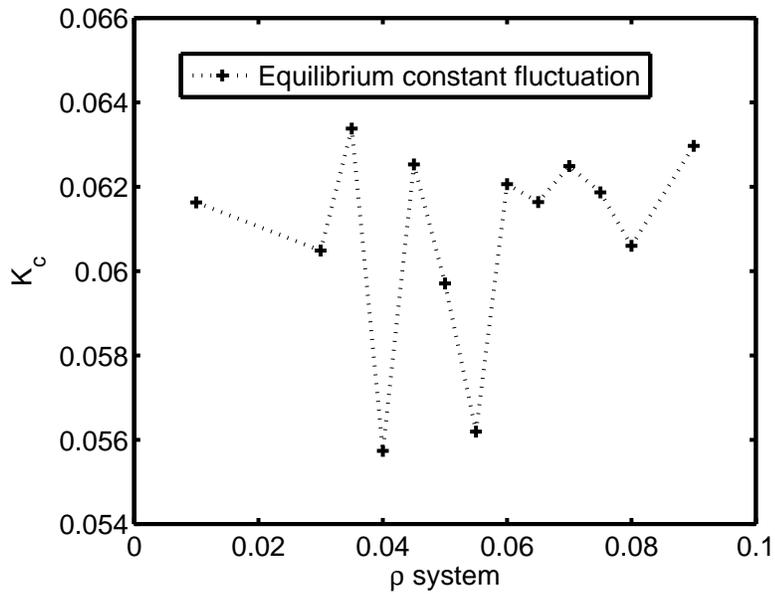} 
\end{center}
\caption{Low density fluctuation of concentration ratio with system density $\rho$. }
\label{fig:8_1} 
\end{figure}
\begin{figure}[htbp]
\begin{center}\includegraphics[%
  width=9cm]{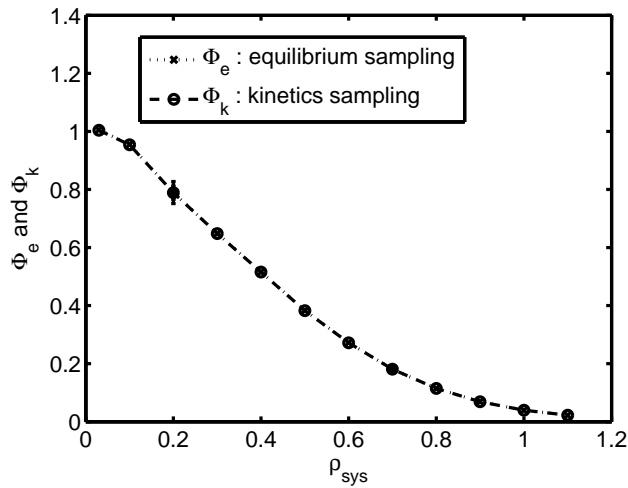} \end{center}
\caption{Variation of $\Phi_e$ and  $\Phi_{k}$ with $\rho$, which is the system
number density at LJ temperature $T^{*}=8.0$ where there is  empirically complete coincidence within experimental uncertainty. }

\label{fig:5_9}
\end{figure}
\begin{figure}[htbp]
\begin{center}
\includegraphics[ width=11cm]{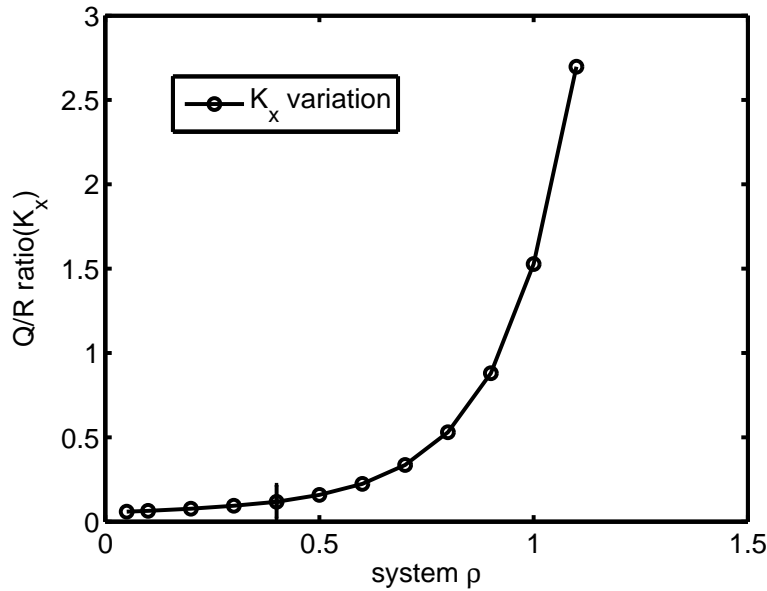} 
\end{center}
\caption{ Variation of $K_x$ with system density, with the  uncertainty of $sde=50$ at $\rho =0.4$.}
\label{fig:4_2} 
\end{figure}

It will be noted that the errors in the $K_c$ ratio  at low densities are of order  20 times than that at higher densities,  where relatively very accurate sampling is possible. What is observed at very low density is a "`saturation"' or limit effect for both rates (leading to fluctuations in $K_x$ in method (ii)  and depicted in Fig.~(\ref{fig:4_2}) and for $K_c$  for this method (i) given in Fig.~(\ref{fig:8_1}). This allows us to determine $K_{eq}$ from taking the average value over a range of low $\rho$ in the saturated range of $\rho$ (here from $\rho =0.03-.09$) over about 13 values where theoretically $K_{eq}=\, \lim M\rightarrow \infty \sum_{i=1}^MK_{c,i}/M$ over any saturated range.
The resulting constant is 
\begin{equation}
K_{eq}=\lim_{\rho\rightarrow0}K_{c}=0.061\pm.002\text{LJ units}\label{kc}\end{equation}
 Knowing this value, we calculate the activity coefficient ratio,
$\Phi_e=\frac{\gamma_{\text{A}_{2}}}{\gamma_{\text{A}}^{2}}$, for the other densities using \begin{equation}
K_{eq}=K_{c}\frac{\gamma_{\text{A}_{2}}}{\gamma_{\text{A}}^{2}}=K_{c}\Phi_e .
\end{equation}
 The ratio of activity coefficients $\Phi_e$ is shown as a function
of density in Fig.(\ref{fig:5_9}).%
It is evident from the plots that the mixture is highly non-ideal, which may be expected due to the
large differences in the LJ energy well for the molecule and the atom
(see Fig.(\ref{fig:1}). 

The separate activity coefficients may perhaps be
derived from cycle studies of equilibrium states that include the
reference state at infinite dilution. The derivation might require
a series of very elaborate and detailed computations. However, to date no clear
theory nor example has been provided  as to how this might be achieved.

(ii) {\itshape Kinetic sampling method} 

\begin{figure}[htbp]
\begin{center}
\includegraphics[ width=11cm]{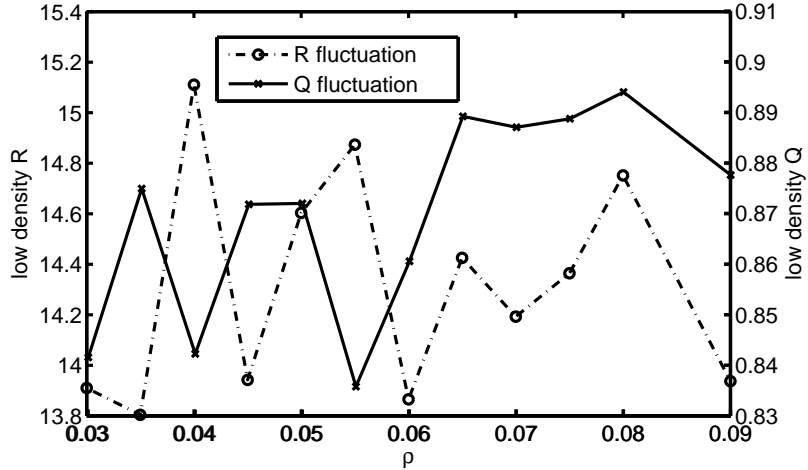} 
\end{center}
\caption{Very low density fluctuation of $R$ and $Q$. These values were averaged to yield the low density limits.}
\label{fig:8_2} 
\end{figure}

\begin{figure}[htbp]
\begin{center}\includegraphics[%
  width=11cm]{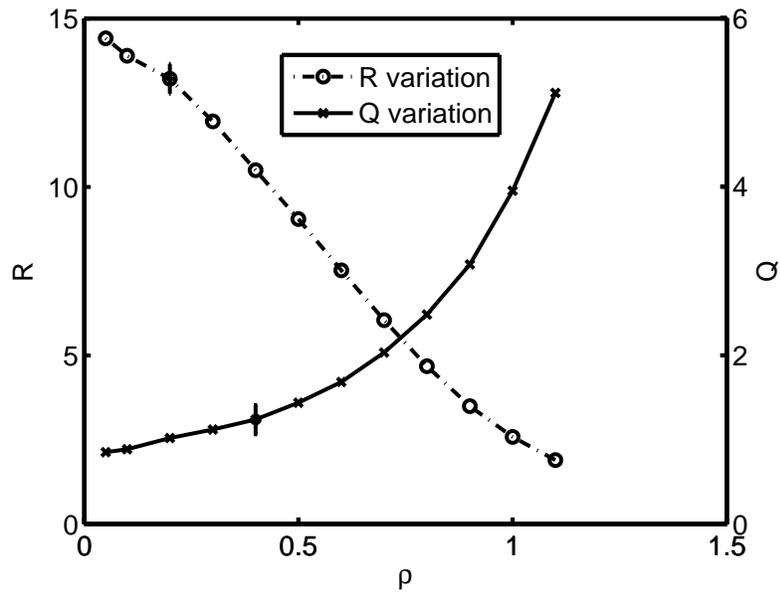} \end{center}
\caption{General variation of $Q=k_{1}$ and $R=k_{-1}$ with $\rho$, the system density at LJ temperature
$T^{*}=8.0$ with $sde (Q)=50$ and $sde (R)=3$ at the indicated points. These values are at typical system  densities which were not used for the extrapolation process.}
\label{fig:7_8} 
\end{figure}

From the way the algorithm was constructed for molecular formation,
the molecularity of the elementary reaction is 2 leading to a single
second-order reaction of formation, and for the dissociation of $\mbox{A}_{2}$, a
first-order reaction results since the molecule can only exchange
kinetic energy with all other particles within the system without
further reactions to the dissociation limit. There are many definitions
and conventions for describing the overall rate of reaction and the individual rate constant. 
Here, 
the forward rate constant $k_{1}^{0}$ and the dissociation rate constant
$k_{-1}^{0}$ are defined as the value of the respective rate constants as $\rho\rightarrow0$
where the unsuperscripted $k$'s refer to the rate constants for non-zero finite 
$\rho$. The overall rate of reaction $r$ may be written in terms
of the experimentally determined forward rate $r_1$ defined as $r_{1}=k_{1}x_{\text{A}}^{2}$ and backward rate $r_{-1}$
defined as $r_{-1}=k_{-1}x_{\text{A}_{2}}$ where
$r=r_{1}-r_{-1}=k_{1}x_{\text{A}}^{2}-k_{-1}x_{\text{A}_{2}}$. \, In this definition of rate, no normalization by the stoichiometric factors of the chemical reaction are used. At
equilibrium $r=0,$ and so 
\begin{equation}
\frac{x_{\text{A}_{2}}}{x_{\text{A}}^{2}}=\frac{k_{1}}{k_{-1}}.
\end{equation}
 The ratio of rate coefficients is the concentration ratio which is this time written as  $K_{x}$ and is expressed only as a kinetic ratio 
where \begin{equation}
K_{x}=\frac{k_{1}}{k_{-1}}.\end{equation}
 To verify the above we plot 
 \begin{eqnarray}
 r_{1}/x_{\text{A}}^{2}&=&Q=k_{1}\nonumber \\
     &\text{and}& \nonumber \\
    r_{-1}/x_{\text{A}_{2}}&=&R=k_{-1}\label{e22} 
 \end{eqnarray}
  against  the density $\rho$ 
and extrapolate to zero density to determine the equilibrium constant.
The rates were calculated independently from the program by monitoring
the number of bonds formed or broken for each time step $\delta t^{*}$
and averaging this quantity over the $10M$ time steps.  The plots of $Q$
and $R$  at low densities  (where saturation is observed) are given in Fig.(\ref{fig:8_2})
As with $K_c$, a saturation effect is observed amidst the evident fluctuations at very low densities and the zero density (superscripted $0$) limit for $Q$ and $R$ is defined thus:
\begin{equation}\label{e:l2}
\begin{gathered}
  Q^0  = k_1^0  = \mathop {M \to \infty }\limits^{\lim } \frac{{\sum\nolimits_{i = 1}^M {Q_i } }}
{M} \hfill \\
  R^0  = k_{ - 1}^0  = \mathop {M \to \infty }\limits^{\lim } \frac{{\sum\nolimits_{i = 1}^M {R_i } }}
{M} .\hfill \\ 
\end{gathered} 
\end{equation}
The subscripted $Q$'s and $R$'s in (\ref{e:l2}) are the values determined in the saturation  region interval of $\rho$ where $0.03\leq \rho \leq 1.1$. In this work $M=13$. If the simulation is to be consistent and at equilibrium, then $K_c=K_x=Q/R$ which is precisely observed in Fig.~(\ref{fig:4_2}). The individual  variation of $R$ and $Q$ are given in Fig.~(\ref{fig:7_8} ).

The results for the low density limits are as follows:
\begin{align}
\lim_{\rho\rightarrow0}Q & =k_{1}^{0}=0.870\pm0.006\text{L.J. units}\\
\lim_{\rho\rightarrow0}R & =k_{-1}^{0}=14.32\pm0.1\text{L.J. units}\end{align}
 and their ratio is \begin{equation}
K_{x}^0=\lim_{\rho\rightarrow0}\frac{k_{1}}{k_{-1}}=0.0610\pm.006\text{L.J. units}\end{equation}
 An excellent agreement with the  equilibrium $K_{eq}$ is found, i.e. $K_{eq}=K_{x}^0=K_c^0$,
where the method used for the determination of
the equilibrium constant differs from the static distributive sampling of method (i). 
\subsection{Survey of some standard form for elementary rate constant form}\label{sub:3.1a}
An elementary reaction has been defined irreducibly \cite[p.1]{ross1} as due to the actual scattering of particles where "`..Chemical reactions occur by the collision of molecules, and such an event is called an elementary reaction for specified reactant and product molecules"'  where if there are two or more elementary steps involved, as in the Michaelis-Menten reaction, then a "`pseudo-elementary reaction"' is implicated \cite[p.3]{ross1}.The rate coefficient \cite[p.7, Chap. 3]{ross1} is written $k(T), \,$ "` and is generally a function of temperature, and frequently of $T$ only."' The onset of perturbations  \cite[eg. Chap. 11, Oscillatory Reactions ]{ross1} are due to species effects not connected to $k(T)$ where the rate $\nu$ of the elementary reaction is written $\nu=k(T)[A]^\alpha[B]^\beta ...$ and the entire field of reaction mechanisms does not question this representation. The form was established over a century and a half ago and follows from the law of mass action \cite[Chap. 3]{kou1} which was suggested by persons which included C.L. Berthollet (1799), M. Berthelot and P. St. Gilles (1862-63), L. Wilhelmy (1850), H. Rose (1842), A.W. Williamson (1859) and C. Guldberg and P. Waage (1864-67) where they all in different ways generally inferred that at equilibrium,
\begin{equation} \label{c1}
k_+R_1R_2....=k_- P_1P_2...
\end{equation}
where the $R$'s and the $P$'s were effectively the mass concentrations of reactants and products and where the $k$'s were "`coefficients of velocity"' which {\itshape were independent of concentration}. This form has been  maintained to the present times no matter how complex the reactions (e.g. possibly involving multi-step cluster and activated thermal electron transfer \cite{tembe1} or complex chemical oscillations \cite{gray1}. In the preface of the text devoted to non-linear chemical kinetics \cite{gray1}, it was surmised "`..only first order processes escape the non-linear net, and even these get caught if there is the slightest departure from isothermal operation."'  On the other hand, in the dimer reaction described here, the first order $A_2 \rightarrow 2A$ process is shown to be non-linear in the most ordinary circumstances. Complex concentration effects on the rate constant have been structured by postulating the concatenation of several elementary reactions, meaning that these reactions are pseudo-elementary according to Ref. \cite{ross1}, where each elementary reaction rate constant is strictly concentration independent in all the references hitherto encountered, and which therefore can be taken to be a universal definition. In Eyring's transition state theory, a scheme
\begin{equation} \label{c2}
Reactants \rightleftharpoons Activated\,\,Complex\rightarrow Products
\end{equation}
is postulated for the  pseudo-elementary reaction (\ref{c2}) above \cite[eqn. 4.3, p.125]{eyr1}, leading to a rate $R$ 
\begin{equation}
	R=k^\prime (T)\frac{\gamma_1\gamma_2}{\gamma^\dag}\ldots C_1C_2\ldots
\end{equation}
where $k^\prime(T)$ is strictly temperature dependent, and the $\gamma$'s are the activity coefficients; $k'(T)$ is derived from a product function involving equilibirum constants, and strictly elementary rate constants, implying no other variable dependency other than $T$. Likewise, pressure dependency of unimolecular and association reactions \cite[1995 edition, p.121-142]{bkreac14} are explained by Lindemann-type mechanisms involving complex equilibria, which introduces a pressure dependence on the pseudo-rate constant \cite[1995 edition, p.138, Sec. 5.11, Association reactions]{bkreac14}. Ref. \cite[Sec. 5.4,p.186 ]{eyr1} give examples of  composite elementary reactions, such as \cite[eqn. 5.90, p.191]{eyr1}
\begin{equation}	
A + M\mathop  \rightleftharpoons \limits_{k_{ - 2} }^{k_2 } A^ *   + M,\;A^ *  \mathop  \to \limits^{k_1 } A^\dag   \to P
\end{equation}
where various quantum statistical approximations are applied to each elementary step involving the individual $k_i(T)$ rate constants to derive the overall rate constant which is first order at low reactant  concentrations and second order at high concentrations. As stated elsewhere, Kosloff has gone beyond the $TST$ pseudo-elementary reaction theories by developing a straight trajectory model \cite[p.187]{kos2} which superimposes trajectories with energy grids. More recent methods, such as developed by Miller  \cite{mil1}involves going beyond TST where an exact trajectory is calculated "`which is no longer a transition state theory"' \cite[p.387]{mil1}. However, in these exact treatments, the isolated participants only are included, e.g. \cite[p.eqn. 46, p.402]{mil1}
in
\begin{equation}
	H + O_2\longrightarrow OH + O
\end{equation}
where the CRP (cumulative reaction probability) is calculated for the 4 atom system ($6-D$ calculations). The microcanonical rate constant $k(T)$ for a bimolecular reaction $A+BC\rightarrow AB+C$ is defined such that $k(T)$ is the canonical ensemble rate coefficient in the expression $-\frac{d[A]}{dt}=k(T)[A][BC]$ and $k(T)$ is determined from an integral involving the CRP. A direct evaluation can also be made without CRP's \cite[p.408, eqn.50]{mil1} where the direct calculation involves a flux operator $\hat{\mathcal F}$ which is related to the system Hamiltonian and trajectory which in all these treatments does not include the rest of the environment variables. A recent example of such direct calculations in given in \cite{arin1} for gas phase reactions, where the flux operator is a type of commutator $\hat{\mathcal F}=\frac{i}{\hbar}[H,\theta (s)]$, where $H$ is the system Hamiltonian and $s$ are the trajectory coordinates with $\theta $ being the heavyside function. Reference \cite{pnas1} is an excellent resume of the most up to date prominent methods available, all conforming to the $k=k(T)$ assumption for elementary reactions where no direct relationship with the activity of the species has been deduced, as will be attempted here. A brief survey of the  recent past also does not yield any exceptions to the standard definition \cite[p.36, p.109, eq. 3.25, p.239, eq. 7.38]{hous1}. Espenson  \cite[p.157]{bkreac15}too, writes $k=k(T)$ for elementary reactions. In developing the standard Br{\"o}nsted-Bjerrum equation for the reaction $A + B \rightarrow P \,(products)$, he breaks down this reaction into  a sum of elementary  reactions \cite[eq.9-22]{bkreac15}
\begin{equation}
	A \,\, + B\,\,\stackrel{K^\dag _a}{\rightleftharpoons}[AB]^\dag\,\rightarrow\,\,P,
\end{equation}

where he infers \cite[p.204]{bkreac15}, in keeping strictly  with the rate constant definition that "`...The rate is taken to be proportional to the concentration (not the activity !) of the transition state"', thereby deducing the standard form \cite[p.204, eq.9-27]{bkreac15} for the Br{\"o}nsted-Bjerrum equations as
\begin{equation}
	k(T)=k_{ref} \frac{\gamma_A\gamma_B}{\gamma^\dag}
\end{equation}
where in the present simulation, the reference state is the vacuum. Other treatises do not differ in interpretation when they write \cite[p.3]{bkreac7}"'...Typically, rate constants are independent of concentration but they may be quite sensitive functions of the temperature."' This independence of concentration for elementary reactions is maintained elsewhere, e.g.  \cite [p.13,eq.1-11]{bkreac1},\cite[p.2]{bkreac3}and \cite[p.35, eq.2.39]{bkreac9}.

\subsection{Reactivity, activity and availability   coefficients and ratios}\label{sub:3.2} 
\begin{figure}[htbp]
\begin{center}\includegraphics[%
  width=10cm]{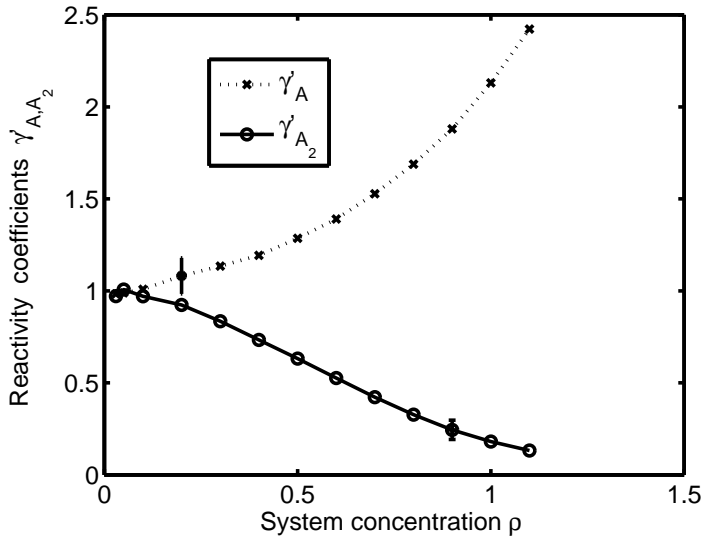} \end{center}
\caption{Variation of reactivity coefficients $\gamma_{A}^{'}\,\,\text{and}\,\,\gamma_{A_{2}}^{'}$
with $\rho$, the system number density at temperature $T^{*}=8.0$. }
\label{fig:10} 
\end{figure}

Write 
\begin{equation}
\begin{gathered}
	  Q = k_1^0 \gamma _A^{'2}  = k_1 \hfill \\
  R = k_{-1}^0 \gamma _{A_2 }^{'}  = k_{-1}  \hfill \\
  \end{gathered}
\end{equation}
where  $\gamma _A^{'}$ and $\gamma _{A_2}^{'}$ are defined as "`reactivity"' coefficients.
In this system, the substrate in which the chemical reaction occurs is the vacuum state with a defined zero of energy relative to the component species. Obviously, $\gamma _A^{'}=\sqrt{\frac{Q}{k_1^0}}
$ and $\gamma _{A_2}^{'}=\frac{R}{k_{-1}^0}$ are computable since all the other terms  are. These coefficients are graphed in Fig.~(\ref{fig:10}).
 We make the following observation:
\begin{obs}
$\gamma _A^{'}$ and $\gamma _{A_2}^{'}$ are not in general unity, especially at higher $\rho$ values.
\end{obs} 
According to  the prevailing theories  over the  centuries, the rate constant proper is independent of concentration for strictly elementary reactions. Activity coefficients are only introduced  via an equilibrium constant as a result of composite schemes where some pre-equilibrium is postulated, the chief example being activated complex theory, and the earlier and related ionic reactions based on the Br{\"o}nsted-Bjerrum equation. In the present scheme, pre-equilibria has been ruled out, because  the non-reversible, cyclical pathway of the reaction  does not allow for such a situation. The results here show that the conventional definition of the rate constant for elementary reactive processes is incomplete, unless the $\gamma^{'}$ reactivity factor is included,  as summarized below.
\begin{lemma}\label{lem:1}
Given that the temperature-only dependent  portion of the rate constant $k_B^0$ is always present, then its product with the reactivity coefficients are a necessary and sufficient condition for the rate constant to be complete.
\end{lemma}
{\bfseries Proof.} Denote by $[X]$ the presence of factor $X$ and $\vee$ $\wedge$ the logical 'or' and 'and' conjunctions respectively with $\neg$ the negation. Let the rate ${\cal S}_B$ for species $B$ be given by 
\begin{equation}\label{e:l3}
{\cal S}_B=k^0_B\gamma^{'}_B
\end{equation}
where $\gamma^{'}_B$ are the reactivity factors (they may be product terms as for our $A$ species). Then $[\gamma^{'}_B]\wedge [k^0_B]\Rightarrow{\cal S}_B$ is complete since form (\ref{e:l3})
 leads experimentally to an exact description for all $\rho$ values. The converse is 
 
\begin{equation} \label{e:l4}
	\neg\left([\gamma^{'}_B] \wedge [k^0_B]\right)=\neg[\gamma^{'}_B]\vee\neg[k^0_B] \Rightarrow\neg{\cal S}_B.
\end{equation}
 Since by hypothesis $[k^0_B]$ is present, then $[\neg k^0_B]$ is false and so if $\gamma^{'}_B$ is not present implies ${\cal S}_B$ is incomplete (which is impossible) $\bullet$
 \newline Fig.~(\ref{fig:10})provides for the following:
 \begin{obs}\label{obs2}
 $\Phi_k=\frac{\gamma_{A_2}^{'}}{\gamma_{A}^{'2}}$ is coincident with $\Phi_e$ within computational  accuracy for all physically feasible $\rho$.
 \end{obs}

This is an remarkable and unexpected  result from which further properties issue forth such as:
\begin{lemma}\label{lem:2}
The general solution to Observation(\ref{obs2})is 
\begin{equation}\label{e:l5}
\begin{gathered}
  \gamma _A^2 c(\rho ) = \gamma _A^{'2}  \hfill \\
  \gamma _{A_2 }^{} c(\rho ) = \gamma _{A_2 }^{'}.  \hfill \\ 
\end{gathered} 
\end{equation}
 where  $\gamma_A$ and $\gamma_{A_2}$ are the activity coefficients.
\end{lemma}
{\bfseries Proof.}The reactivity coefficients for any species $X$ is known i.e. $\gamma^{'}_X=\gamma^{'}_X(\rho)$ and $\gamma_A,\gamma_{A_2}$ exists by hypothesis (Observation (\ref{obs2}). Then $\exists$ $d_i$ such that $\gamma _A^2d_i=\gamma _A^{'2}$  and similarly    $\gamma _{A_2 }d_i^{'}=\gamma _{A_2 }^{'}$.Observation(\ref{obs2}) implies $\left(\frac{d_i^{'}}{d_i}=1\right)$
for all $\rho$ or $d_i^{'}=d_i$ for every $\rho_i$. Define the function $c:\rho_i\rightarrow d_i$ or $c=c(\rho)\,\bullet$

If $c(\rho)$ is a more complex function than unity, we may rescale our apparent
activity coefficients such that we may relate the kinetically derived
$\gamma^{'}$ reactivity coefficients with the 'actual' $\gamma$ coefficients determined
from equilibrium distribution studies, so that $\gamma_{A}^{'}=\gamma_{A}c(\rho)^{1/2},\gamma_{A_{2}}^{'}=\gamma_{A_{2}}c(\rho)$ are used instead of the $\gamma$ activity coefficient in the chemical potentials for the system. That is, if the actual potentials are scaled, we wish to determine whether $\gamma_{A}^{'}$ and
$\gamma_{A_{2}}^{'}$ may be used as equivalent activity coefficients when $\gamma_A$ and $\gamma_{A_2}$ are scaled
by functions $c(\rho)^{1/2}$ and $c(\rho)$ respectively. As a matter of interest,  in the sections that follow, it will be shown that the chemical  potentials refer to purely isothermal reversible work terms.
 \begin{lemma}\label{lem:3}
\label{lm1} The scaled activity coefficients yield equivalent expressions
for the Gibbs equilibrium criterion for the same Gibbs standard chemical
potentials, but the Gibbs-Duhem condition demands $c(\rho)=\mbox{constant}$.
\end{lemma} 
\textbf{Proof.} Writing the rescaled chemical potential
for any species $p$ as $\mu_{p}^{+}=\mu^{*}+RT\ln x_{p}\gamma_{p}^{'}$
and $\mu_{p}=\mu^{*}+RT\ln x_{p}\gamma_{p}$ for the unscaled potentials,
then application of Gibbs' criterion for the scaled variables yields:
$2\mu_{A}^{+}=\mu_{A_{2}}^{+}\Rightarrow2\mu_{A}=\mu_{A_{2}}$ by
algebraic cancellation and $2\mu_{A}=\mu_{A_{2}}\Rightarrow2\mu_{A}^{+}=\mu_{A_{2}}^{+}$
by algebraic addition of $\ln c$ on both sides of the equation. Hence
either set ($\mu$ or $\mu^{+}$) of potentials may be used to determine
the equilibrium point where $\gamma_{p}^{'}$ represents the activity
coefficient for the {$\mu_{p}^{+}$} chemical potential set. The
Gibbs-Duhem equation for the scaled system at constant ($P,V$) is
\begin{equation}
\sum_{i}x_{i}d\mu_{i}^{+}=0\label{a1}\end{equation}
 where

\begin{equation}
d\mu_{p}^{+}=d\mu_{p}^{*}+RTd\ln x_{p}+RTd\ln\gamma_{p}+RTd\ln f_{p}(c).\end{equation}
 Comparing (\ref{a1}) with $\sum_{i}x_{i}d\mu_{i}=0$ for the $\{\mu_{p}\}$
set, we derive \begin{eqnarray}
x_{A}RTd\ln c^{1/2} & + & x_{A_{2}}RTd\ln c=0\\
 & \mbox{or} \nonumber \\
\frac{x_{A}}{2}d\ln c=x_{A_{2}}d\ln c\end{eqnarray}
 which is generally a contradiction unless $c(\rho)$=constant, which implies
$c=1$ \, since $\lim_{\rho\rightarrow0}c(\rho)=1\,\bullet$

 \begin{coro}
\label{coro1} From Lemma (\ref{lem:3}), either set ($\mu\,\mbox{or}\,\mu^{+}$)
may be used to determined the equilibrium point, but there is only
one specified activity coefficient set \{$\gamma_{p}$\}. 
\end{coro}

\begin{coro} \label{coro2} If it can be proved that for any rescaled
activity coefficient $\gamma_{i}'$, there can exist only one equilibrium
concentration ratio satisfying the Gibbs equilibrium criterion and
the Gibbs-Duhem equation for the same chemical potential functions,
then $c(\rho)=1$ and $\gamma_j^{'}=\gamma_j$. \end{coro} 


\begin{coro}\label{coro4}
The activity coefficients are unique in that if they are rescaled, then $c(\rho)=1$ only.
\end{coro}

\begin{coro} \label{coro5}
From the invariance
of the equilibrium constant due to the Gibbs criterion, the rate constants
between two states 1 and 2 due to catalytic activity are related as
follows: \end{coro} 

\begin{equation}
\left(\frac{k_{1}}{k{-1}}\right)_{1}\left(\frac{k_{-1}}{k_{1}}\right)_{2}\left(\frac{\gamma_{A_{2}}}{\gamma_{A}^{2}}\right)_{1}\left(\frac{\gamma_{A}^{2}}{\gamma_{A_{2}}}\right)_{2}=1\label{e28}\end{equation}
 One may not expect the $\gamma$'s to vary between two equilibrium
catalytic states by mutual cancellation but eq.(\ref{e28}) is 
the general expression. Given the same mechanism, changes of rate
can only be affected by changes in the isolated intermolecular potentials,
or the net potentials due to the sum of all relevant interactions. Both
these cases seem to imply minute changes to the species type. Provided
that these changes do not affect to first order $\Delta G_{r}^{0}$
for the standard free energy change, eq.(\ref{e28}) should obtain
approximately. Since $Q,R,\text{and the }\gamma's$ are functions
of $\rho$, we can write equations (\ref{e22}) in the form

\begin{eqnarray}
Q & = & k_{1}^{0}\left(\sum_{i=1}^{\infty}a_{i}\gamma_{A}^{i}\right)\gamma_{A}^{2}\nonumber \\
R & = & k_{-1}^{0}\left(\sum_{i=1}^{\infty}b_{i}\gamma_{A_{2}}^{i}\right)\gamma_{A_{2}}\end{eqnarray}
 If $R$ and $Q$ are independent variables, it is not immediately
obvious why the common factor $c(\rho)=\sum_{i=1}^{\infty}a_{i}\gamma_{A}^{i}=\sum_{i=1}^{\infty}b_{i}\gamma_{A_{2}}^{i}$
must have the same non-constant value at each $\rho$ if we should
vary the temperature: hence one might expect that $c(\rho)=1$: if
this is not generally the case, then there is embedded within the
MD potentials a yet to be clarified link between the forward and backward
rates.

\begin{figure}[htbp]

\begin{center}\includegraphics[%
  width=10cm]{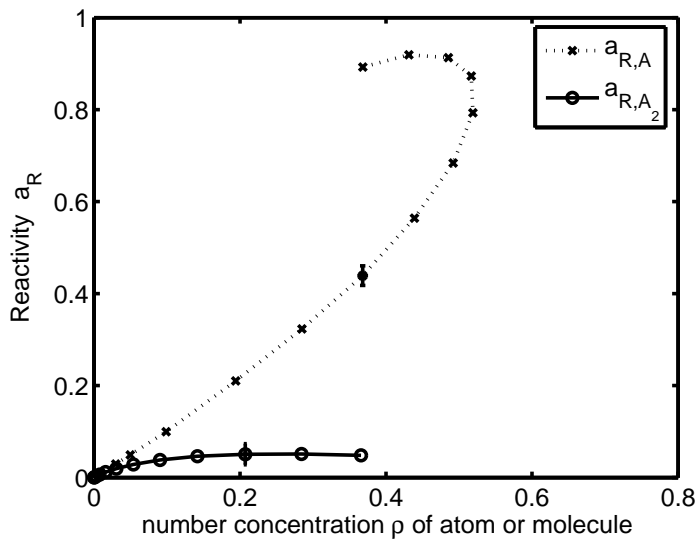} \end{center}

\caption{Variation of reactivity $a_{A}$ and $a_{A_{2}}$ with $\rho_{A}$ and $\rho_{A_{2}}$,
the actual number density of the species concerned at LJ temperature
$T^{*}=8.0$ with $sde=10, 40$ respectively for the uncertainty in $a_{A}$ and  $a_{A_{2}}.$ }

\label{fig:11} 
\end{figure}

The plot of the reactivity coefficient is given  in Fig.~(\ref{fig:10})as
a function of average system density. Both coefficients extrapolate
to unity. While the reactivity coefficient of A is larger than unity,
that of  $A_{2}$ is smaller than unity, and one
might  expect such behavior to arise from the differing potentials and
net surface area to volume ratio that exists for these two species.
The reactivity of species $p,a_{p}$ is defined  by $a_{p}=\gamma'_{p}.\rho_{p}$
where $\rho_{p}$ is the concentration (e.g. number density) of species $p$. A plot of
the variation of these quantities appears in Fig.~(\ref{fig:11}),
where the densities refer to the actual number density of the concerned
species and not to the general system density $\rho$. The looping
of the curve for $\text{A}$ is very interesting because with
increasing system density, the equilibrium constant adjusts itself
to accommodate an increasing reactivity coefficient for $\text{A}$,
implying at constant temperature a species density reduction, which
explains the looping back behavior of the reactivity for $\text{A}$. At the current state of development, one must use physical arguments to determine whether the reactivity and activity coefficients are identical or not. It is surmised that it is not from considerations of the  estimates of the LJ fluid activity coefficients relative to the boundary   conditions imposed. It is estimated that  both the dimer and atomic  activity  coefficients either increase or decrease with system density, and therefore cannot equal the reactivity coefficients where the trend for dimer and atom are in contrary directions. It is then surmised  that the activities both increase with system concentration by comparison with the LJ fluid for the atomic constituent. Before examining the thermodynamics of the single component monomer A, we digress to an outline of a theory of internal species equilibrium that is of use to describe species states along a reaction pathway.

\subsubsection {Internal equilibrium  of product species}\label{subsub:3.3.0}
It was conjectured  that a subsystem described by non-canonical coordinates would have a Boltzmannized energy distribution for the kinetic energy if the energy interaction of the coordinate were due solely to external forces (due to  the reservoir). This conjecture is extended to the external potential here; in fact the Debye-Huckel (DH)  and extended theories of electrolyte solutions all make this assumption  \cite{bkelectro1}. Under this assumption, a particular species $Z$ (being either product $P$ or reactant $R$ species in the equilibrium $R \rightleftharpoons P$)  can exist in  $p$ varying  energy states $Z_1,Z_2,\ldots Z_p$ as enumerated by an appropriate  algorithm. The enumeration may be with respect to energy states in terms of a coarse grid of magnitude $\Delta E$; such techniques have been developed in recent times by Kosloff  \cite[p.187]{kos2}who superimposes reaction dynamics trajectories  with a gridded "`energy range of molecular encounter $\Delta E$ ...typically in the range of $0 - 10$ eV."'  And It will be further assumed that a convenient  algorithm exists for this purpose. The example given here is for reasons of illustration, to show that the activity coefficient may be determined at any point if a partial enumeration were known, and to extend the DH theoretical assumption. The science of enumeration has made much progress in recent years, especially in the characterization of species types in mathematical chemistry, whose pioneers include Balaban\cite{lyg1}. The techniques developed for chemical structures could conceivably be extended to these cluster types.  The temperature of  each of the species in the enumeration is approximated as the same as the system temperature $T$ determined from $\frac{3kT}{2}=\left\langle \frac{p_i'^2}{2m_i}\right\rangle$ where it has been shown that they may differ slightly  since they are not the canonical coordinates \cite{cgj10}. Hence this theory is an approximation. Then the chemical potential $\mu_{Z_i}$ of each of these states share the same standard state $\mu^0_Z$ where $\mu_{Z_i}=\mu^0_Z+kT\ln c_i \gamma_i$ with $c_i$ and $\gamma_i$ representing the concentration and activity coefficient respectively and for the $Z$ species as a whole, its concentration $c_{tot}$ will be $c_{tot}=\sum_{i=1}^pc_i$ and the bulk activity coefficient will be denoted as $\gamma_{tot}$.  The species equilibrium 
\begin{equation} \label{eg1}
Z_1\rightleftharpoons Z_2\rightleftharpoons Z_3 \ldots\rightleftharpoons Z_p
\end{equation}
implies a form
\begin{equation} \label{eg2}
kT\ln c_{gm}\gamma_{gm}=kT\ln c_i \gamma_i=kT\ln c_j\gamma_j\,\,(j\neq i)
\end{equation}
where $c_{gm}$ and $\gamma_{gm}$ will be determined. From (\ref{eg2}), let $\alpha_i=c_i\gamma_i$, then $\alpha_i=\alpha_j=\alpha'$ since $\mu_{Z_i}^0(T)=\mu^0_Z(T)$ , and further define 
\begin{equation} \label{eg3}
kT\ln\left\{\prod_{i=1}^p c_i\gamma_i\right\}=kT\ln A .
\end{equation}
From (\ref{eg3}),
\[
kT p\ln \alpha'=kT\ln (c_1c_2\ldots c_p) +kT\ln (\gamma_1\gamma_2\ldots \gamma_p)
\]
leading to 
\begin{equation} \label{eg5}
kT \ln \alpha'=kT\ln \{c_1c_2\ldots c_p\}^{1/p} +kT\ln \{\gamma_1\gamma_2\ldots \gamma_p\}^{1/p}
\end{equation}
Define 
\begin{eqnarray}
	\ln c_{gm}&=&\ln \{c_1c_2c_3\ldots c_p\}^{1/p} \nonumber \\
	\ln \gamma_{gm}&=&\ln \{\gamma_1\gamma_2\gamma_3\ldots c_p\}^{1/p}.	
\end{eqnarray}
Then $c_{gm}=\{c_1c_2\ldots c_p\}^{1/p}$ and $\gamma_{gm}=\{\gamma_1\gamma_2\ldots \gamma_p\}^{1/p}$. Further (\ref{eg5})gives 
\begin{equation} \label{eg7}
kT\ln c_{gm}\gamma_{gm}=kT\ln\alpha'=kT\ln\gamma_i c_i.
\end{equation}
It will be noted that (\ref{eg7}) can refer to a partial (incomplete) enumeration where the $i^{th}$ enumeration need not be in the set if the Gibbs equilibrium criterion obtains for the $i^{th}$ species in equilibrium with the other species.
\newline
What is the relationship between the different $c$ and $\gamma$ coefficients?
\newline
For a complete enumeration over an energy grid, each of which has an energy difference   of  magnitude   $\Delta E$, the average potential experienced by the species $X$ is $\overline{w}$, so that $\overline{w}=kT\ln \gamma_{tot}$ by definition, where  $\mu_{Z}=\mu^0_Z+kT\ln c_{tot} \gamma_{tot}$. The probability of state $i,\, p_i$ is $p_i=c_i/c_{tot}$ and under the  the canonical  ensemble assumption above, $p_i=\frac{\exp -\epsilon_i \beta}{\mathcal{Z}}$  where ${\mathcal Z}=\sum_{i=1}^p \exp -\epsilon_i \beta$ is the partition function of the distribution with $\beta=1/(kT)$.
By taking averages, the above yields
\begin{equation} \label{eg12}
kT\ln \gamma_{tot}=kT\sum_{i=1}^p p_i\ln\gamma_i
\end{equation}
or 
\begin{equation} \label{eg13}
\ln \gamma_{tot}=\sum_{i=1}^p \frac{c_i}{c_{tot}}\ln \gamma_i
\end{equation}
Hence, (\ref{eg13}) leads to
\begin{equation} \label{eg14}
c_{tot}\ln \gamma_{tot}=\sum_{i=1}^p c_i\ln \gamma_i
\end{equation}
Defining a departure from bulk value $\gamma'$ where $\gamma_i=\gamma_{tot}\gamma'_i$, then (\ref{eg14}) can be written in the form 
\begin{equation} \label{eg15}
\sum_{i=1}^p c_i\ln \gamma'_i=0.
\end{equation}
by expansion of terms.

If the degenerate terms are all collected together,  the canonical distribution can be written $p_i=\frac{g_i\exp-\beta\epsilon_i}{\mathcal{Z}}$. (Other modifications are straightforward and here $\mathcal{Z}=\sum^p_{i=1} g_i \exp-\beta \epsilon_i$). We assume $p$ non-degenerate energy states. Since $\gamma_i=\exp \beta \epsilon_i$, then the  probability term is $p_i=\frac{g_i}{\gamma_i \mathcal{Z}}$ and so
\begin{equation} \label{eg16b}
\ln p_i=\ln g_i -\ln \mathcal{Z}-\ln \gamma_i
\end{equation}
so that (\ref{eg13}) becomes 
\begin{equation} \label{eg18}
\ln \gamma_{tot}=\sum^p_{i=1} \frac{g_i}{\gamma_i \mathcal{Z}}\ln \gamma_i
=\frac{1}{ \mathcal{Z}}\sum^p_{i=1} g_i \frac{\ln \gamma_i}{\gamma_i }
\end{equation}
and for $g_i=1$ there results 
\begin{equation} \label{eg22}
\ln \gamma_{tot}=\frac{1}{\mathcal{Z}}\ln \prod^p_{i=1}\left(\gamma_i^{1/\gamma_i}\right).
\end{equation}
What can be said of $\gamma_{tot}$ in terms of  $c_{gm}\gamma_{gm}=\alpha$ where  (\ref{eg7}) gives $c_{gm}\gamma_{gm}=\alpha=\gamma_i c_i$? From these definitions and (\ref{eg14}), there results

\begin{equation} \label{eg23}
\ln \gamma_{tot}=\frac{c_{gm}\gamma_{gm}}{c_{tot}}\ln \prod^p_{i=1}\left(\gamma_i^{1/\gamma_i}\right).
\end{equation}
and comparing (\ref{eg23}) with (\ref{eg22}) implies
\begin{equation} \label{eg24}
\mathcal{Z}=\frac{c_{tot}}{c_{gm}\gamma_{gm}}.
\end{equation}
For degenerate systems, let $a_i=g_i/\gamma_i$, then 
\begin{equation} \label{eg25}
\ln \gamma_{tot}=\frac{1}{\mathcal{Z}}\ln \prod^p_{i=1}\left(\gamma_i^{a_i}\right).
\end{equation}
Then (\ref{eg23}) and (\ref{eg25}) leads to a general coupled equation
\begin{equation} \label{eg26}
\prod^p \gamma_i^{a_i/\mathcal{Z}}=\prod^p_{i=1} \gamma_i^{\left(\frac{c_{gm}\gamma_{gm}}{c_{tot}\gamma_i}\right)}.
\end{equation}
Forcing independence of the $\gamma_i$ is equivalent to equating  exponents, leading to the result
\begin{equation} \label{eg27}
\frac{g_i}{\mathcal{Z}}=\frac{c_{gm}\gamma_{gm}}{c_{tot}}.
\end{equation} 

Similarly, (\ref{eg14}) and the definition of $\alpha$ leads to
\begin{equation} \label{eg28}
\ln\gamma_{tot}=\ln\alpha -\sum^p_{i=1} p_i \ln c_i.
\end{equation}
Using the definition for $c_i$ in terms of $p_i$, we derive 
\begin{equation} \label{eg29}
\ln\gamma_{tot}=\ln\frac{\alpha}{c_{tot}} -\sum^p_{i=1} p_i \ln p_i
\end{equation}
 which has an interesting entropy-like contribution in the probabilities.
 Finally, the maximum available potential energy for any one species $\gamma_i$ must obey elementary relations such as $\gamma_{gm}c_{gm}=\gamma_i c_i$ and $\frac{\gamma_i c_i}{\gamma_j c_j}=1$, implying that one can determine the $\gamma$ terms if the concentration terms were known at any volume element of the chemial trajectory.  
\subsection{Thermodynamics of fluid and  discussion of the $c(\rho)$ scaling function}\label{sub:3.3}
Lemma(\ref{lem:3}) implies that since the activity coefficients are unique relative to scaling, an attempt to estimate $c(\rho)$ can come only from physical/kinetic experimental considerations,where this function should not contradict ancillary data and estimates. One can expect the activity coefficient of the $\text A$ atom to be approximately that of the LJ fluid at the same density and temperature at low dimer concentrations. A rationale must be provided on how to estimate this quantity from available data. Some data exists \cite{gub1,gub2} for the LJ fluid up to $T^\ast=6.0$. At such temperatures, the fitting is highly sensitive to the $\gamma$ coefficients, unlike what was reported \cite[sec.4,p.607]{gub2} (presumably for much lower T values, where $\gamma$ was regressed for $1\leq \gamma \leq 7$ with low correlation of the minimum). The parameters for the various estimates for the monomer activity was scanned from the $x(i)$ values in Table 4. of \cite{gub1} and fitted into the equations found in \cite{gub2}. The residual or excess Helmholtz free energy in reduced units, which is the difference between the ideal gas value and the actual value   is given by \cite[eq. 5]{gub2}
\begin{equation} 
	A_r^\ast=\sum_{i=1}^8 \frac{a_i\rho^{\ast i}}{i}+\sum_{i=1}^6 b_iG_i
\end{equation}
 with the pressure having the expression 
 \begin{equation}\label{a2}
	P^\ast=\rho^\ast T^\ast + \sum_{i=1}^8 a_i\rho^{\ast(i+1)}+F\sum_{i=1}^6 b_i\rho^{\ast(2i+1)}
\end{equation}
 with $F=\exp(-\gamma\rho^{\ast 2})$, $\gamma$ being  a non-linear adjustable parameter. 
The excess free energy $G^\ast_r$ is given by 
\begin{equation} \label{a3}
G^\ast_r=	A_r^\ast+\frac{P^\ast}{\rho^\ast}-T^\ast .
\end{equation}
The reduced units variables are denoted by $\ast$ superscripts, and the $a_i$ and $b_i$ coefficients are given as complex polynomials in $x(i)$ and $T^\ast$ in the tables in \cite{gub2}.
 It was found that at $T^\ast=6$ , a good value for $\gamma$ (to the nearest digit) was $2.7$ which yielded  the fitting pressure of $12.7931$, close to the experimental value of $12.43(2)$ in the phase diagram. The simulations here were off-scale at $T^\ast=8.0$, but nevertheless the value of $\gamma$   given above was adopted for estimating the excess free energies at the simulation temperature as well as the atomic $\gamma_A$ coefficients according to the theoretical model below. 

A circular reasoning has been detected in the definition of multicomponent Gibbsian free energies \cite[p.706, Sec. 5]{cgj9} and the associated chemical potential, where it was pointed out that the $TdS$ term defined by Denbigh \cite{den1} sometimes referred to a heat input term for multicomponent closed systems, and sometimes not, leading to an undecided paradox, but where the chemical potential is a pure work term; in the new exact thermodynamics \cite[p.707, after eqn(30)]{cgj9}, the chemical potential is a mixed heat-work term. In this work, we shall conform to conventional descriptions, thereby reinterpreting the physical meaning of some of the standard expressions. It will be argued here that relative to the concentration where the chemical potential is written $\mu_i=\mu^0_i +kT\ln c_i\gamma_i$, and where this term represents isothermal work done on the system, the excess Helmholtz energy 	$A_r^\ast$ seems to refer to the $\gamma_i$ activity  coefficient (here analogous to the "`fugacity"') coefficient and not the excess Gibbs free energy $G^\ast_r$ for the reasons that follow. For a single phase at constant temperature, $dG=Vdp$ and for a perfect classical fluid $\Delta G =\int Vdp=-\int pdV$ where the work loss on the system is equivalent to $\Delta G$ but this is {\itshape not true} of imperfect systems. It will be proved from the Kelvin-Clausius statement that the external work (if it were the sole work source) takes into account the internal  intermolecular forces. 
The chemical potential $\mu_i\equiv\left(\frac{\partial G}{\partial n_i}\right)_{T,P,n_j}$, despite being defined as \cite[p.79]{den1} "`......the amount by which the capacity of the phase for doing work (other than work of expansion) is increased per unit amount of substance added...."' is often written in the form 
\begin{equation} \label{e:l6}
\mu_i=\mu_i^0(T) + kT\ln c_i\gamma_i
\end{equation}
 for substance $i$. The definition by Denbigh may perhaps be reflected in the view of some others who suppose that even for  single phase systems, it is believed that $G/N\equiv \mu_i$ \cite[pg 83]{lands1}. Indeed, for a single phase system, 
 some have identified the chemical potential with the Gibbs function as $U-TS+pV=\mu n=G$ with n being the amount parameter \cite[Sec. 7.3,p.83]{lands1}.

  This would suggest that for multicomponent systems, the free energy differential  written 
\begin{equation}\label{e:l7}
dG=-SdT +VdP + \sum_i \mu_idn_i
\end{equation}
 implies that the chemical potential refers to a contribution to the free energy to the state at constant $p$ and $T$ \cite[p.126]{ira1} with the material addition $dn_i$. Then from $G=\sum_in_i\mu_i$ \cite[p.255,eqn.9.23]{ira1} , the one component form suggests that $\frac{\partial G}{\partial n_i}=\mu_i$ or $\frac{\partial (\mu_i n_i)}{\partial n_i}=\mu_i$ or 
 $\frac{\partial (kT\ln c_i \gamma_i)}{\partial n_i}=0$ for all species $i$ which is undemonstrated. Furthermore, for imperfect fluids $\int V dP\neq -\int PdV$ generally and so does not appear to constitute a pure external work term for single phase, single component systems; if it were nevertheless the case, then 
 \begin{eqnarray}
  G_{ex,i}=G_{real,i}-G_{id,i}&=&\mu^0(T)+kT\ln c_i\gamma_i-\mu^0(T)-kT\ln c_i        \nonumber \\
  \Rightarrow \gamma_i&=&\exp\left(\frac{G_{ex,i}}{kT}\right).\label{g1}
\end{eqnarray}

To estimate the activity coefficient of the LJ fluid, one must refer to the excess work done on the particle $\delta w_{ex,i}$, where conventionally, the activity coefficient $\gamma_i$  refers to this work. \cite[ Chap.1,p. 4-8]{bkelectro1} where $kT\ln \gamma_i =\delta w_{ex,i}$; it will be shown that the term $kT\ln c_i$ refers  to the external work done  on the system  to the state concerned, and this work is also equivalent to introducing unit amount of substance $i$  from the standard state condition of unit activity coefficient. The work done for a perfect fluid on the system is 
\begin{equation}\label{e:23}
\delta w = -NkT\ln\frac{V_2}{V_1}=NkT\ln\frac{c_2}{c_1}
\end{equation}
between states $2$ and $1$. Under the typical convention $c_1\rightarrow0,\,\gamma_1\rightarrow 1$, the standard state $c_1$ leads to a singularity in this limit. Write $\delta c_i M_i=1$ for any small value of $c_1$ of amount $\delta c_1$ when $\gamma_i\rightarrow 1$. So, $M_i \rightarrow \infty$ but for any small $\delta c_i$, it is a large finite number. The form of the chemical potential, if rescaled to conform to these limits must be of the form
\begin{equation}\label{e:24}
\mu^{\prime}_i=\mu^{0\prime}_i(T)+kT\ln \frac{c_i\gamma_i}{[1Unit]}.
\end{equation}
But whatever the scaling, the form and value for these variables match, i.e. $\mu_i=\mu^\prime_i$ where $\mu_i^0$ refers to a standard chemical potential of isolated substance $i$ as $c_i\rightarrow 0$ since the physical meaning of $\mu_i$ must be preserved. Thus,
\begin{equation}\label{e:25}
\mu^{0}_i(T)+kT\ln \frac{\gamma_ic_i}{[\delta c_i]}=\mu^{0}_i(T)+kT\ln M_i+kT\ln \frac{c_i\gamma_i}{[\delta c_i M_i]}
\end{equation}
so that 
\begin{equation}\label{e:26}
\mu^{0\prime}_i(T)=\mu^{0}_i(T)+T.Q_i
\end{equation}
where 
\begin{equation}\label{e:27}
Q_i=k\ln M_i.
\end{equation}
Let substance $i$,  also termed substance $A$ be the basic constituent from which all other substances are formed. Other substances are built upon this primary substance, where in general 
\begin{equation}\label{e:28}
A_n+A_m \rightarrow A_{m+n}
\end{equation}
and the subscript $m$ refers to the number of units of $A$ making up substance $A_m$. For instance for the reaction $nA\rightleftharpoons A_n$, the standard state chemical potential  $\mu^0_{A_n}(T)$ for $A_n$ is defined as 
\begin{equation}\label{e:29}
\mu^0_{A_n}(T)=n\mu^0_{A} +\delta\mu^0_{A_n}.
\end{equation}
Here,
$\delta\mu^0_{A_n}$ refers to the work required to form $A_n$ from $n$ units of $A$. For all reacting species $A_q$, a common lower end concentration limit is imposed where 
\begin{equation}\label{e:30}
\delta c_{A_n}=\delta c_{i}=\delta c_{A}=\delta c_{A_q}
\end{equation}
for all $n$ and $q$. Because of this, we can write 
\begin{equation}\label{e:31}
\mu^{0\prime}_{A_n}(T)=n.\mu^{0\prime}_{A} +\delta{A_n}
\end{equation}
where 
\begin{equation}\label{e:32}
\mu^{0\prime}_{A}=\mu^0_i(T) + TQ_i 
\end{equation}
with $Q_i=k\ln M_i$.
But from (\ref{e:32}) 
\begin{equation}\label{e:33}
n\mu^{0\prime}_{A}=n\mu^0_i(T) + nTQ_i 
\end{equation}
and from (\ref{e:31}) the following results

\begin{eqnarray}
    \mu^{0\prime}_{A_n}   & = &n\mu^{0}_{i}(T) + nTQ_i +\delta A_n \label{e:34}\\
       & = &n\mu^{0}_{i}(T) + nTQ_{A_n} +\delta A_n \label{e:35}
 \end{eqnarray}
But (\ref{e:29}) yields by comparison to (\ref{e:35})

\begin{equation}\label{e:36}
 \mu^{0\prime}_{A_n}= \mu^{0}_{A_n}+ nTQ_{A_n}.
\end{equation}
The equilibrium criterion gives
\begin{equation}\label{e:37}
-\Delta G^0=-\Delta G^{0\prime}=-\sum_{j=1}^M\mu_j^0\nu_j=-\sum_{j=1}^M\mu_j^{0\prime}\nu_j=kT\ln K_e
\end{equation}
and (\ref{e:36}) gives
\begin{equation}\label{e:38}
\sum_{j=1}^M\mu_j^{0\prime}\nu_j=\sum_{j=1}^M\mu_j^{0}\nu_j+\sum_{j=1}^M\left(n_jTQ\right)\nu_j
\end{equation}

Now , $n_j$ is the number of elementary species $i$ that constitutes $j$, and this quantity is always conserved (from a chemical point of view) and $T$ and $Q$ are common, leading to $TQ\sum_j n_j\nu_j=0$. This verifies (\ref{e:37}) and infer that the same equilibrium point is reached. It follows that we may ignore the terms that cancel off if only equilibrium problems are of significance, and write the chemical potential for equilibrium problems as
\begin{equation}\label{e:39}
\mu^{\prime\prime}_i(T)=\mu^{0}_i(T)+kT\ln \frac{c_i\gamma_i}{[1]}
\end{equation}
where the potential here is a measure of the work done on a particle $i$ in isolation where the singularity of the point at $\infty$ ($c_i=0$)has been removed. The above establishes the fact that the form $\mu^{\prime\prime}_i(T)=\mu^{0}_i(T)+kT\ln c_i\gamma_i$ 	is an isothermal work term with removed singularities at zero concentration, consisting of the work used to  overcome the internal forces and the external ideal work.
\subsubsection{On Helmholtz heat-work interchange}\label{subsub:3.3.1}
 An approximate value for the multicomponent activity coefficient may be derived from the Helmholtz free energy for single systems by considering the work-heat transformations. The single component Helmholtz free energy $A$ is given in standard form  as $A=U-TS$, where  its differential $dA=dw-SdT$ implies that at constant $T$ , the work done on the system $dw$ is the so-called external work (such as the $P-V$ work connected with compression) where for fluids, $dA=-PdV=dw$. For fluids, 3 basic categories ((i)-(iii))of compression from state at concentration $\delta c_1$ to $c$ (where the lower concentration is at $\delta c_1$)  must be examined  as follows:
\newline (i){\itshape ideal fluid compression.} 
   
The reversible work of compression is $\delta w =kT \ln \frac{c}{\delta c_1}$ and let the heat absorbed be $-\delta q$. The energy change between the states (subscripts denote the state for the variable concerned) $\Delta E_i=E_c -E_{\delta c_1}=\delta w - \delta q$.
\newline (ii){\itshape compression of gas by by manipulation of internal forces.} 
Here two separate sub-parts are considered where (a) is generally applicable and is in  the standard format of a conventional thermodynamical system and (b) develops an internal potential description with control of the particle potential. For both cases, divide the path into interval segments with states $S$ indicated by subscripts : $ S_0 (\text{at}\,\delta c_1), S_1,S_2 \ldots S_n (\text{at}\,c)$. For (b),we suppose that the external pressure $P$ of the imperfect gas  obeys $P\geq P_{id}$, due to the repulsive intermolecular forces relative to the perfect gas with pressure $P_{id}$ for any concentration $c_i$ for state $S_i$. For a segment for which the opposite  inequality holds, the same conclusion  of the theorem obtains and so a general path may be broken up into subsections with one or other of the inequalities obtaining (where it is assumed that the pressure is continuous over the path), implying the general validity of the theorem. 
 Method (a): Start with  a perfect gas and compress reversibly to $S_n$. At $S_n$, we "`charge up"' the gas by introducing a potential with an interaction parameter $\epsilon$  say. In Debye-Huckel theory of solution activity, where the charge of the ions assumes the variable $\epsilon$; in a LJ type potential system, a possible $\epsilon$ parameter  for the potential $\mu (r)$ is $\mu(r)=\epsilon \left(\frac{1}{r^6}-\frac{1}{r^{12}}\right)$. For any potential and given particle coordinates, we can in principle compute the total potential of interaction. For each particular value of $\epsilon$ we can calculate the mean energy of interaction per unit increment of $\epsilon$. Integrating this yields the average energy of potential interaction as $\overline{w_{f_{a}}}$. During this charging up process, heat energy will be absorbed by the system of amount $-\delta q^\prime_m$  during the manipulation where the system is at a fixed temperature  $T$. For the entire compression process, the total heat absorbed is 
 \begin{equation} \label{e:40}
 -\delta q^\prime=-\delta q^\prime_m -\delta q
\end{equation}
Then the total change in energy is 
\begin{equation} \label{e:41}
\Delta E(ii)=\delta w + \overline{w_{f_{a}}}-\delta q^\prime
\end{equation}
where $\overline{w_{f_{a}}}$ is  perhaps a novel diathermal work term involving internal potential variables; the standard methods  hitherto used seem to suppose no heat transfer during the charging process and possible fundamental flaws in the  thermodynamics might be introduced as a result. 
\newline (iii)The third process is reversible isothermal compression of the imperfect gas by doing $P-V$ work from $S_0$ to $S_n$ at concentration $c$,where $-\delta q_c$ of heat is absorbed by the system. The total change of energy  will be given by
\begin{equation} \label{e:42}
\Delta E(iii) =\delta w_c -\delta q_c.
\end{equation} 
Since the final and initial states are the same, then the First Law implies $\Delta E(ii)=\Delta E(iii)$ but nothing thus far can be said about the $\delta q$'s, i.e. $\delta q^\prime$ need not equal $\delta q_c$. Suppose in fact that the heats are not the same $\delta q^\prime \neq \delta q_c$. A cycle of state transitions can be carried out as follows:
\begin{equation} \label{e:43}
S_{\delta c _1}\stackrel{step(ii)}{\rightarrow}S_{c}\stackrel{step(iii)}{\rightarrow}S_{\delta c _1}
\end{equation}
For this cycle, the  First Law yields
\begin{equation} \label{e:44}
\delta w+ \overline{w_{f_{a}}} -\delta q^\prime - \delta w_c + \delta q_c=0
\end{equation} 
leading to 
\begin{equation} \label{e:45}
(\delta w+ \overline{w_{f_{a}}}  - \delta w_c )=(\delta q^\prime-\delta q_c)\neq 0
\end{equation}
or the  effect in this cycle is the \underline{NET} conversion of heat to work about an isothermal cycle, which is in contradiction  to the Kelvin-Clausius Second Law postulate \cite[p.101]{kes1}. Hence, $\delta q^\prime = \delta q_c \,\,\bullet$. 
\newline For the above, we have extended the scope of the traditional understanding of the system as given to manipulation of the internal potential of force interactions.  This result may be stated thus:
\begin{theorem}\label{t1}
The heat absorbed by a system  during an isothermal transition is equivalent to that which is absorbed by an ideal system and the heat absorbed when an intermolecular potential is introduced within the system. 
\end{theorem}
\begin{coro}\label{c5}
The reversible work performed during a system transition is that due to the work for an equivalent ideal system and the potential energy change due to intermolecular forces and therefore the excess work done relative to the ideal fluid is due to the intermolecular potential energy  change of the system.
\end{coro}
 {\itshape Proof.} The first part follows from  $\delta w_c =\delta w + \overline{w_{f_{a}}}$ from (\ref{e:45}) and theorem \, \ref{t1}, and the second from $\delta w_c -\delta w =\overline{w_{f_{a}}}\,\bullet$
\begin{coro}\label{c6}
The change in the excess Helmholtz free energy $\delta A_{ex}$is equal to the change in the intermolecular potential energy for the system.
\end{coro}
{\itshape Proof.} By definition, $\delta A_{ex}=\delta w_c-\delta w$ and this quantity is given by  corollary \ref{c5} by $\overline{w_{f_{a}}}\bullet$
\begin{coro} \label{c7}
The activity coefficient for a single component system $\gamma_i$  relative to the standard state is given by $\gamma_i = \exp \left(\frac{\delta A_{ex}}{kT}\right)$.
\end{coro}
{\itshape Proof.} Since $\delta w_c=kT\ln \frac{\gamma_i c}{\delta c_1}$ ,  $\delta w=kT\ln \frac{c}{\delta c_1}$, and $\delta A_{ex}=\delta w_c-\delta w$ then $\delta A_{ex}=kT\ln \gamma_i \bullet$

\begin{figure}[htbp]
\begin{center}
\includegraphics[ width=11cm]{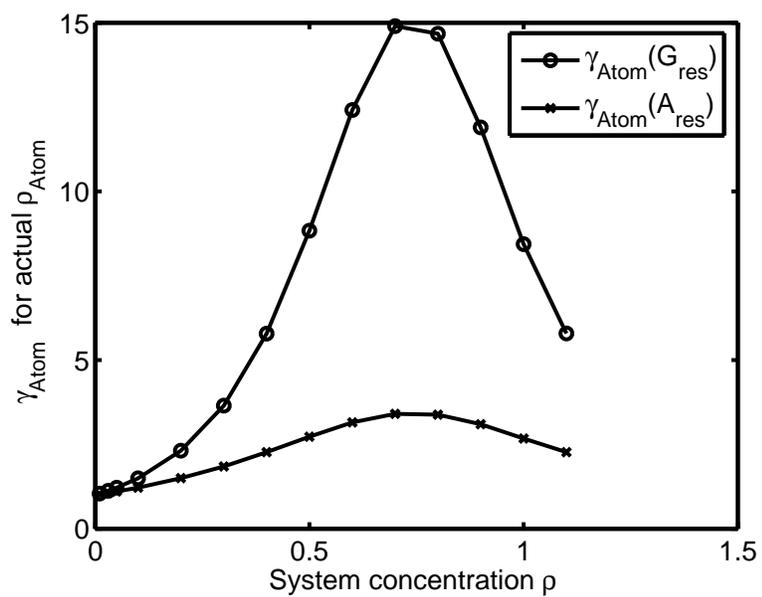} 
\end{center}
\caption{Computed monomer-only activity from residue Helmholtz ($A$)and Gibbs ($G$) functions }
\label{fig:15} 
\end{figure}

\begin{figure}[htbp]
\begin{center}
\includegraphics[ width=11cm]{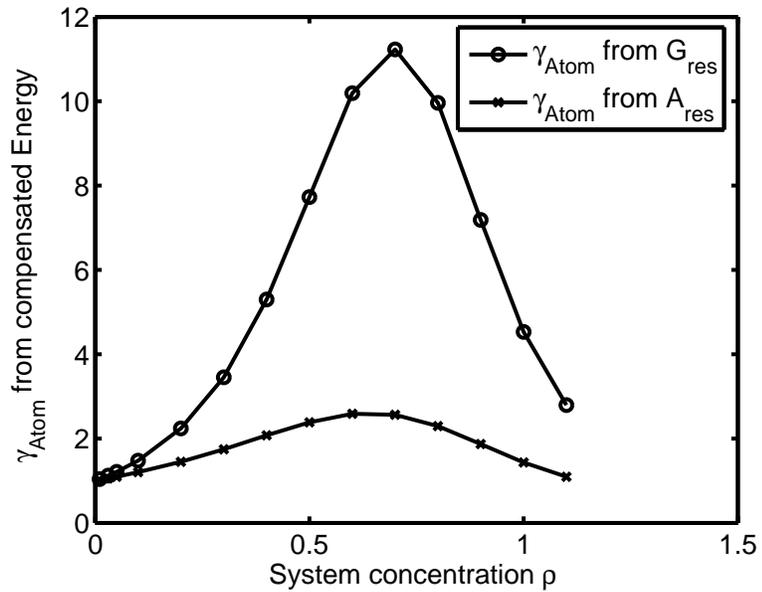} 
\end{center}
\caption{Estimated activities from the residual energy left after dimer formation using density distribution data from the simulations.}
\label{fig:16} 
\end{figure}

In the dimeric reaction $2\text{A}\rightleftharpoons \text{A}_2$, for lower $\text{A}_2$ concentrations (lower $\rho$ values), the limit $\gamma_{A,\text{pure}}$ (pure phase) $\rightarrow \gamma_{A,\text{reaction}}$, the activity coefficient for the reactive system A is exactly as for the single component fluid because the force fields  acting on A are essentially  the same as for the pure component, except there would be clumps of A atoms held together by intermolecular harmonic potential whenever $\text{A}_2$ is present, but these internal forces do not affect the external forces acting on the A atom. This identification  and limit is  used to choose the general form of $\gamma \,vs.\, \rho$ curve for atom and dimer, since the two  different assumptions that are used leads in one case to semi-quantitative agreement of  $\gamma_A$ for the reaction and $\gamma_{A,\text{pure}}$ (pure phase), and it is therefore inferred that the system conforms better to one of the two assumptions made to derive estimates of the activity coefficients.
Clearly, only one of these must obtain from the previous deduction (Corollary \ref{coro4}) concerning the uniqueness of the activity coefficient. We can infer that the results from Fig.(\ref{fig:12}) is the more consistent one  by referring to the monomer situation, for which some data exists, and by estimating the monomer activity from the data and to then infer that the form and trend provided for the monomer at lower concentrations should be approximately that for our pure atomic species as deduced from readily available functions. These functions were estimated for values of temperature up to $T^\ast=6.0$ and therefore are not adequate for higher temperature estimations. Nevertheless. even for $T^\ast=6.0$, the values follow exactly the same trends as for the results here at $T^\ast=8.0$, including the activity estimates all much greater than unity. 
\begin{figure}[htbp]
\begin{center}
\includegraphics[ width=11cm]{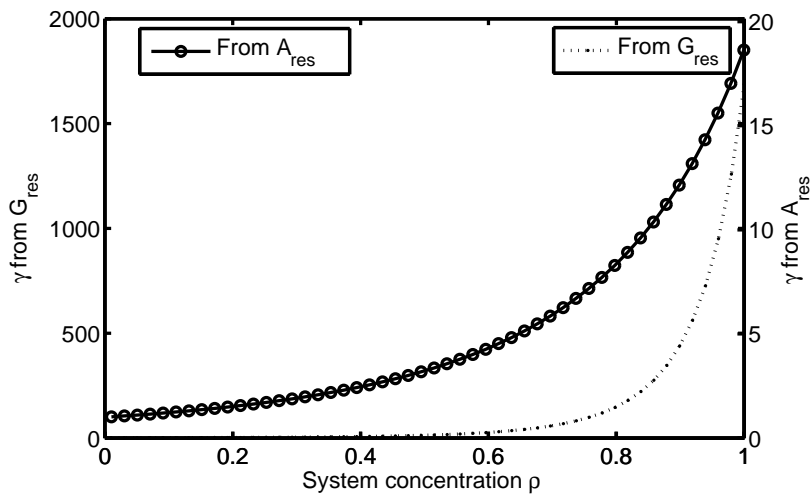} 
\end{center}
\caption{Computed monomer-only activity from residue Helmholtz ($A$)and Gibbs ($G$) functions }
\label{fig:14} 
\end{figure}
The $\gamma$ estimates are derived from eqn.(\ref{g1}) and the expression from Corollary(\ref{c7}). It is clear from Fig.(\ref{fig:14}) that the activity estimates show values all greater than unity. The system $\rho$ for this figure and all others refer to $\rho_{system}$, all in reduced units where $\rho_{system}=\rho_A+2\rho_{A_2}$, where $\rho_A$ is the number density for the monomer (Atom) and $\rho_{A_2}$ refers to the dimer species number density in this homogeneous system at equilibrium. Fig.(\ref{fig:14}) is computed for the situation $\rho_{Atom}=\rho_{system}$.Since what is being depicted is the external potential work done in bringing the atom to the system, then the forces acting on the atom can arise  from either the dimer or the atoms within the system, and since the potentials are of the same LJ form for each nucleus, then we would expect this density to be the most appropriate one. If we just consider what the activity might be for the actual density of the monomer, ignoring its interaction with the dimer, then then  Fig.(\ref{fig:15}) gives the results from the simulation  results and the fitting functions for $G_{res}$ and $A_{res}$ where the system concentration $\rho\equiv \rho_A$.  
Finally, if we supposed  that whatever energy that is available to the monomer is lost whenever a dimer is formed from the activation energy, then the available energy per atom is given by Fig.(\ref{fig:16}) . What is meant in this case is that let the residual free energy per particle be $F_r$, where $F_r=G_{res}$ or $F_r=A_{res}$.The vacuum activation energy is 17.652. So the net energy left per particle is surmised to be $F_{res}=\frac{F_rN_A-17.652N_{A_2} }{N_A}=F_r-17.652\rho_{A_2} /\rho$ and $\gamma_{F_{res}}=\exp(F_{res}/(kT)^\ast)$ which is a measure of the available energy after dimer formation. This expression is plotted in the figure mentioned above where $N_A$ is the number  of atoms for that general system concentration in the MD cell. These figures are plotted to determine which would be the best approximation to the first order theory of elementary reactions developed here.
\subsection{Considerations in Hamiltonian systems}
Recently, it has been proved \cite{cgj12} that the Liouville theorem,and in particular the Liouville space,  of cardinal  importance to both classical \cite{han1} and quantum-mechanical theories of dynamics and statistical  mechanics \cite{pen1} does not obtain  typically in standard applications; and therefore that non-Hamiltonian theories posited using Liouville space, such as developed by Tuckerman et al. are equally suspect \cite[ Tuckerman et al. references within]{cgj12}. A work which does not imply a steady state but a disintegrating system which preserves a certain form  of entropy had been described earlier without recourse to Liouville space and operators, and which considers the Jacobian of a transformation \cite{cgj14}. "'Non-Hamiltonian"' systems has been variously described, including their relation to thermostatting using a pseudo-Hamiltonian that forces constant conditions such as pressure or temperature \cite{frenk1}. It should be noted that many of the synthetic thermostats were designed with the belief in the Liouville theorem. As such, it would be expected that equilibrium properties might be computed if the phase-space covered follows the canonical energy distribution for the coordinates. However, if the cause of the canonical distribution is not due to a smooth Hamiltonian trajectory, but to external random impulses, then it can be inferred that the real time behavior   of particles described by such pseudo- Hamiltonians over any  short enough time interval $\delta t$ would not follow the prescribed pseudo-Hamiltonian trajectory. Such an inference implies that the dynamical behavior of the ensemble of particles would not correspond to the actual possible trajectories with random perturbing forces. It then means that molecular simulation with such  pseudo-Hamiltonians would not yield realistic trajectories which could explain real-time, non-averaged behavior over very short time intervals where random impulsive forces were acting; in particular, biological system function based on real time changes of coordinates  cannot be accurately simulated by such pseudo-Hamiltonians if their main mode of dynamics involves random perturbations,such as from a reservoir. Another presupposition concerns  the static nature of a Hamiltonian system \cite[p.496]{frenk1} and the supposed effects of "`walls"' and constraints in the Hamiltonian: "`...The notion in phase space of Hamiltonian system is similar to that of an incompressible liquid; in time the volume of the "`liquid"' does not change. In contrast, a non-Hamiltonian system is compressible. This compressibility must be taken into account when considering the generalization of the Liouville equation to non-Hamiltonian systems"'. 
On the other hand, from the work presented in \cite{cgj12},it  may be inferred that for any Hamiltonian form 
\begin{equation}\label{he1}
	\mathcal{H}_E=\sum \frac{p_i^2}{2m}+V(r)
\end{equation}
one can consider the "`walls"' and other physical constraints of the system as perturbations to $\mathcal{H}_E$ where $V(r)$ does not refer to the wall interactions; for perfectly reflecting walls, for instance, a micro-canonical distribution results, the canonical one obtains for the {\itshape single} system. The system trajectory is Hamiltonian over the time intervals  $\tau_i=t_i-t_{i-1}$ for the set of time coordinates $\left\{t_0,t_1,t_2....t_m,...t_\infty \right\}$ of interference by random energy interchange by the wall or/and thermostat  at time $t_i$; here the walls and energy interchange achieves at least two functions (i) it defines the boundary $\partial C$ and (ii)acts as a thermostat in terms of energy interchange and the change of system trajectory not determined by the Hamiltonian. If the energy levels are quasi-continuous -- which certainly is for this dimer system-- then $\mathcal{H}_E$ will be distributed with a canonical energy distribution $P(E)\propto \exp -\beta\mathcal{H}_E$  according to standard statistical mechanical theories \cite[p.1-3]{feyn1}. For the Hamiltonian system given here, eqs.(\ref{e3late}-\ref{eq:5}), the K.E. profile of   (i) the "`Atom"' is confined to that segment of the potential curve where there is no bonding, and (ii) the product $A_2$ to the molecular potential well, with (iii) the "`transition state"' point located at two different points $r_f$ and $r_b$. The coordinates of the defined quantities (i-iii) are not in general the canonical coordinates of the entire system Hamiltonian. These coordinates represent  transient species  and  the energy terms for these coordinates may or may not be canonical \cite{cgj10}, and if they have a Canonical energy (CE) distribution- such as the K.E. of the centre-of-mass of the dimer- then its apparent temperature defined as $\frac{3kT_a}{2}=<K.E._{c.m.}>$ need not correspond to $T_s$, the system temperature. It was discovered \cite{cgj10} that the mean K.E. of the atom A and dimer about the C.M. followed the C.E. distribution, with apparent temperatures slightly differing from $T_s$. This result is not unknown, since within a reacting system, apparently different temperatures have been experimentally detected \cite{huang1} for the different species in a plasma. Gibbs and his equilibrium criterion suggests that all species at equilibrium must have the same temperature; we observe that experimentally this is not the case, where relative to the C.M., the temperature corresponding to the K.E. is close but not equal, i.e.
$T_{A}\approx T_s,T_{A_2}\approx T_s$ but $T_A \neq T_{A_2}$. Currently, it is unclear whether $T_A=T_{A_2}=T_s$ as $\rho \rightarrow 0$ but this assumption was used for a new theory of energy interconversion \cite{cgj7}. For an elementary reaction species $i\rightarrow \,products$ at very low pressure or concentration, the rate constant $k_i^0(T)\,\,(\rho\rightarrow 0)$ ,  is derived  assuming that the work done against the interparticle force due to reactant  $A_i$ is derived from the kinetic energy of the particles at a particular geometrical locus determined by the impact parameter and geometry of the potentials and the activation energy; all these factors $\Omega$ are almost always considered invariant in Q.M./classical reaction rate theories since they are strictly mechanical properties, so that $k^0_i=k^0_i(\Omega,T_{i,o})$  where it is conventionally assumed that $T_s=T_{i,o}$; for this model, the minimum impact parameter is at $r_f=0.85$ with the activation energy at $17.5153$. At other concentrations, the reaction pathway system would have to be perturbed; this situation is not normally considered even in advanced treatments \cite{pnas1}; the perturbation suggested here are the energy levels of activation; there would be other effects, of first and higher orders which perturbation theory can provide. The object of the current work is to provide an outline of a theory with estimates of the more important effects which can account  for Observation(\ref{obs2}) and Lemma(\ref{lem:2}).  A second order effect concerns the change of mean kinetic energy of the molecules from the zero state $\delta W_{i,K.E.}$ (meaning the state as the density $\rho\rightarrow 0$) where 
\begin{equation}
\delta W_{i,K.E.}	= K.E.(CM,T_{i,\rho})-K.E.(CM,T_{i,o})
\end{equation}
but would not be considered further; obviously this perturbation would affect the number of activated atoms/molecules. The general perturbation is from a reaction at  zero density  (almost always  the form given in standard theories)  with rate constant $k^0_i=k^0_i(\Omega,T_{i,o})$ to that at any other density which is the actual equilibrium state (a.e.s.)  $k_i=k_i(\Omega,T_{i}(T_s))$; i.e.  a mapping from these two states is required. In the ideal state formulation (for example SCT), absolutely no other environmental interaction $\phi_e$ of particle/complex reactants $i$ with $j$  is envisaged up to the collision distance $\sigma$. In the perturbed state, on the other hand,  for $r_{ij}> \sigma$, $\phi_e(r_{x})\neq 0,x={i,j}$ for the coordinate $r_x$ of particle $x$. We note that for  the vacuum state for which $k^0_i$ is defined, $\phi_e=0$ for any particle participating in the reaction.  The work done $\delta w_{i,\rho} $ in transferring a particle $i$  from the standard state of zero density  (identified with the rate constant $k^0_i$) to a.e.s. is trivially 
\begin{equation} \label{he2}
\delta w_{i,\rho} =kT\ln \gamma_i
\end{equation} 
with $\gamma_i$ being the activity coefficient. The concentration term corresponding to the ideal external work does not feature since it is a common term for both the ideal and non-ideal states. In (\ref{he2}), $\gamma$ can be determined from MD by calculating the mean potential $\overline{\phi_i}$for any  system $\rho$, where $\gamma_i=\exp \frac{\overline{\phi_i}}{kT}$. Standard MD gives the same quantity by a particle insertion method  \cite[p.349, Appendix C]{Hal7.1}  where $\mu_{res}=kT\ln\gamma_i$ and $\mu_{res}$  has the form 
\begin{equation} \label{he3}
\mu_{res}=-kT\ln \left[\frac{1}{<kT_{in}>^{3/2}}\left\langle \left(kT_{in}\right)^{3/2}\exp\left(\frac{-U_t}{kT_{in}}\right)\right\rangle\right]
\end{equation}
with $U_t$ being the instantaneous potential of the test particle interacting with other particles.
An obvious generalization of eq.(\ref{he1}) involves writing a Hamiltonian of $M$ primary particles with a boundary $\partial C$ not incorporated into the Hamiltonian in terms of a potential, with  instantaneous energy $E_i$, which would fluctuate when it exchanges energy with the boundary which incorporates a heat reservoir in the form 
\begin{equation}\label{he4}
	\mathcal {H}_{E_i,\partial C}=\sum_i^M \frac{p_i^2}{2m_i}+V(r_1,r_2,....r_i....r_M).
\end{equation}
Although the Hamiltonian (\ref{he4}) is invariant in its form, the system $\partial C$ determines the density of energy levels and the relative energy levels; by equipartition, the average kinetic energy is invariant for all enclosed boundaries, but the potential energy is not, since it is possible to set $\partial C$ such that the intermolecular distance of particles $i$ and $j$ is $\left|r_{ij}\right|\leq C$ and I could therefore arrange a lower bound by suitable compression and choice of the boundary  $\partial C$ such that $V(r_1,r_2,....r_i....r_M)> B_{\partial C}$. The Canonical distribution for the Hamiltonian would obtain if the degeneracy is very great, according to  the principles developed in the Feynman development\cite{feyn1}. Thus relative to the zero density state, where the maximum $\left|r_{ij}\right|\rightarrow\infty$ and $min \left|V(r_1,r_2,....r_i....r_M)\right|=0$, the mean potential energy changes with the boundary (and therefore system density); further, the mean potential experienced by  dimer and reactant atoms would also differ with density changes; hence relative to the zero density reaction scheme, the additional potential energy present is available to overcome the activation barrier of the reaction. The total available energy is  $\delta w_{i,\rho}$, and for the ground state reactants, this energy would lower the apparent activation energy, and will be used to parameterize the change in activation energy; it will be noted that there might exist the possibility  that not all this energy might be utilized in extremely fast reactions where "`molecular chaos"' and the adjustment of neighboring particles is not rapid enough for the full utilization of this potential. Further investigations, both theoretical and experimental would be required to elucidate these remarks. Hence a flexible approach is required to interpret the $\gamma$'s which may in some cases be a fraction $f$ in value to the thermodynamical $\gamma$; the results here indicate $f\approx 1$.

\begin{figure}[htbp]
\begin{center}
\includegraphics[ width=9cm]{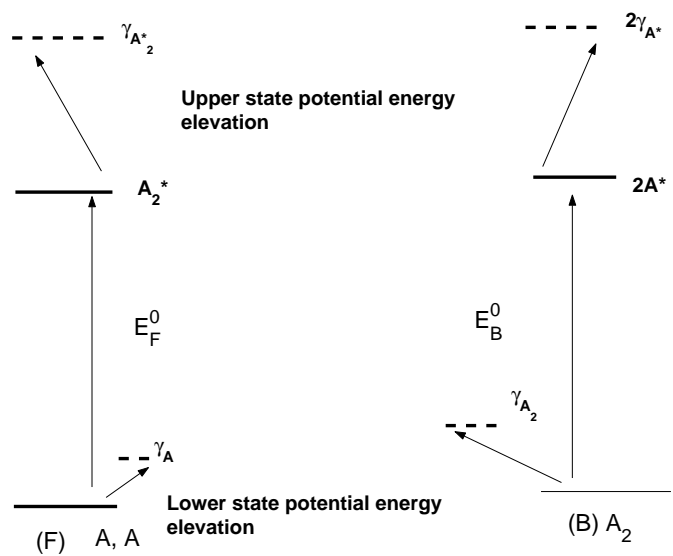} 
\end{center}
\caption{Single stage model with external potential interaction leading to chemical reaction, and which is similar  to the elementary Bj{\o}rn-Bjerrum theory of ionic reactions where (B) denotes the backward direction and (F) the forward direction of the Dimer reaction in eq.(\ref{e1}).}
\label{fig:d2} 
\end{figure}

\begin{figure}[htbp]
\begin{center}
\includegraphics[ width=9cm]{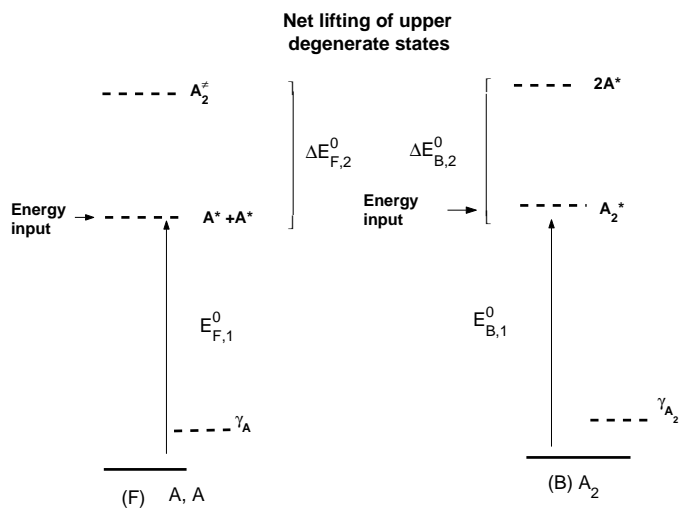} 
\end{center}
\caption{Double  stage model with external potential interaction leading to chemical reaction, which goes beyond the single stage model and analogs where (B) denotes the backward direction and (F) the forward direction of the Dimer reaction in eq.(\ref{e1}).}
\label{fig:d1} 
\end{figure}

  The proposed models for particle interactions yielding chemical reactions for the forward (F) and backward (B) steps are given in Figs.~(\ref{fig:d2}-\ref{fig:d1})below.These models enable one to calculate the activity coefficients where one model contradicts experimental observation whereas the other does not, in addition to corroborating the deductions of the activity from thermodynamical functions.
  
\begin{figure}[htbp]
\begin{center}
\includegraphics[ width=11cm]{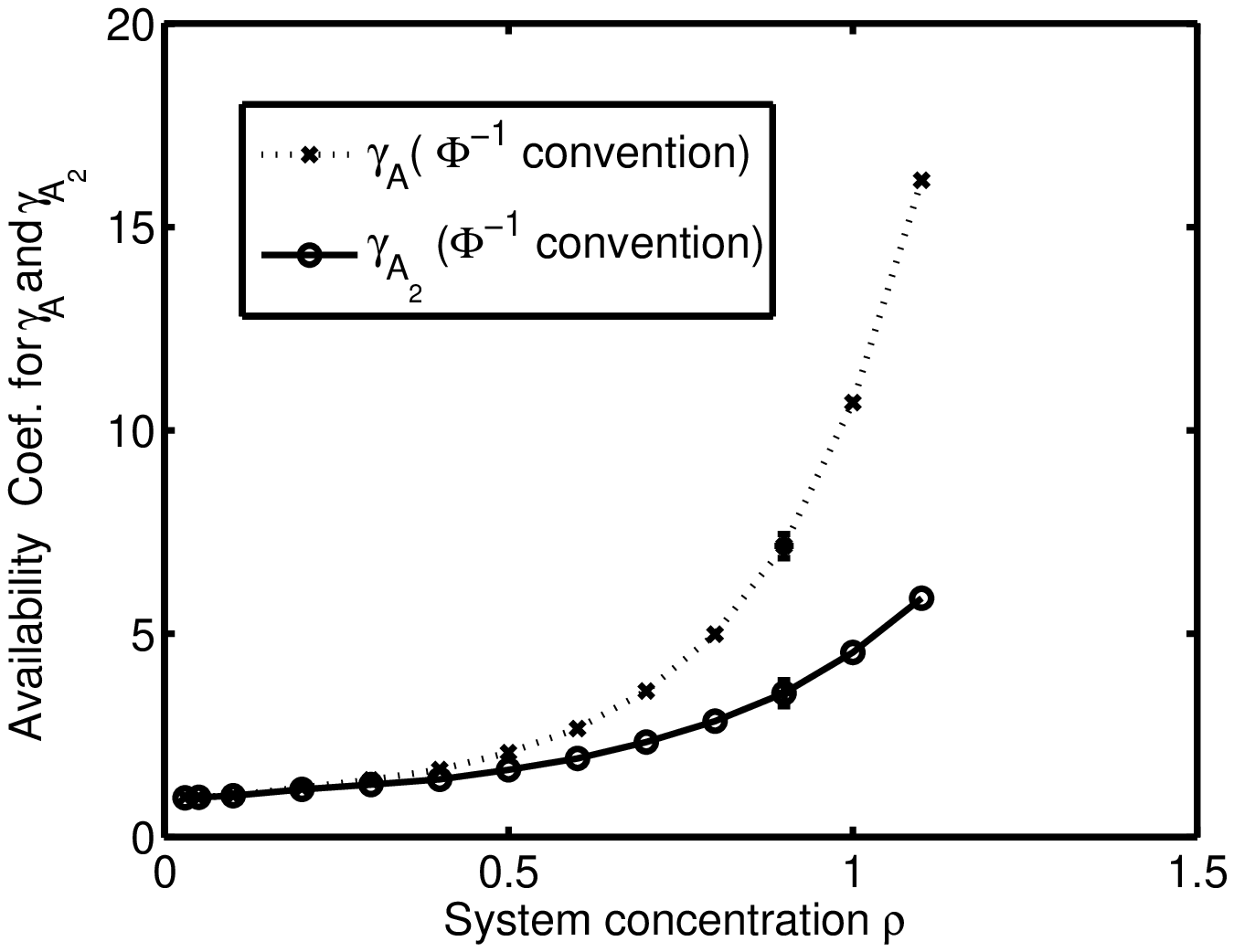} 
\end{center}
\caption{Activity coefficients using the inverse $\Phi$ convention due to boundary conditions.}
\label{fig:12} 
\end{figure}

\begin{figure}[htbp]
\begin{center}
\includegraphics[ width=11cm]{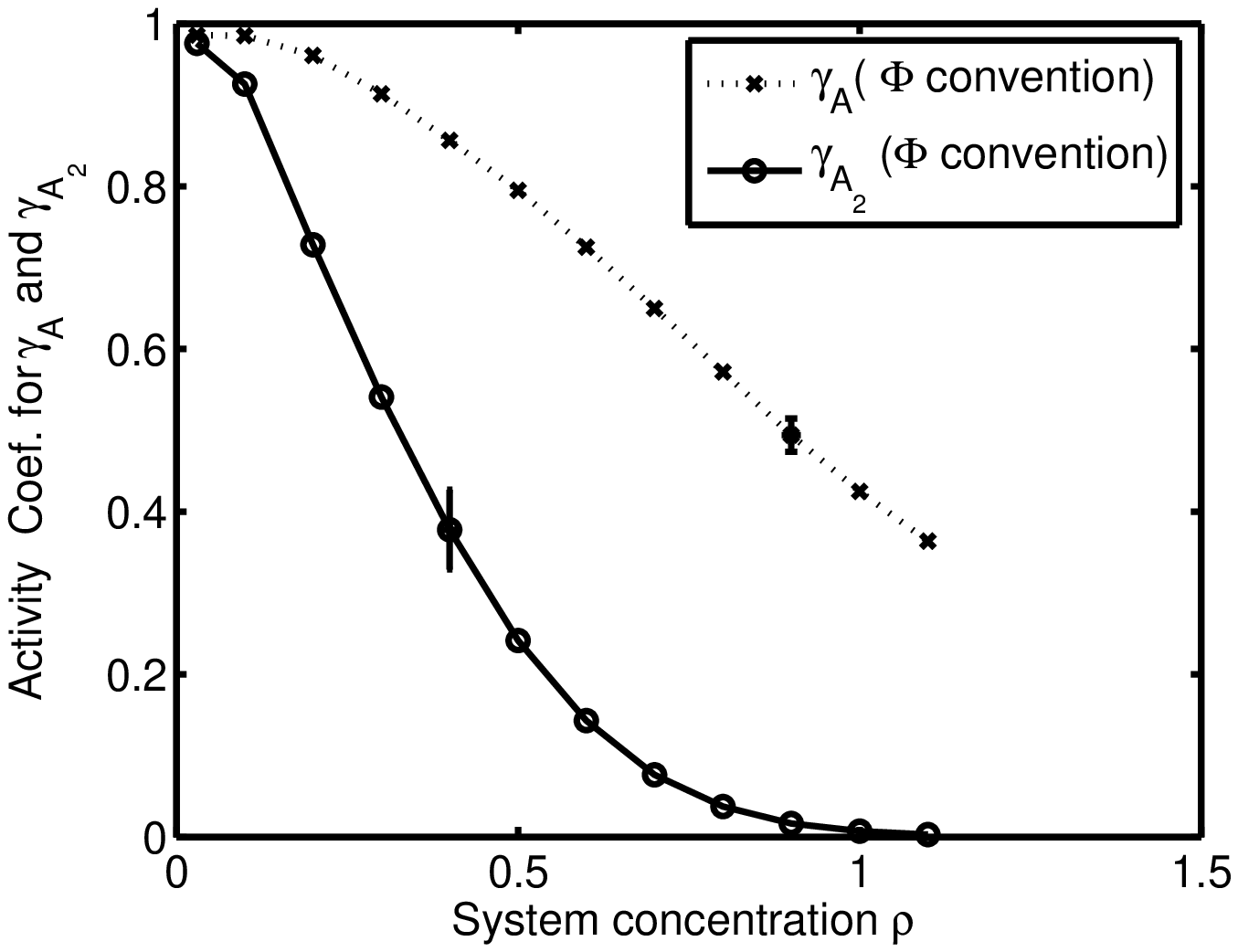} 
\end{center}
\caption{$\Phi$ convention determination of activity coefficients. }
\label{fig:13} 
\end{figure}

The inverse $\Phi$ convention yields the activity coefficients given in Fig.(\ref{fig:12}). On the other hand, the non-inverse $\Phi$ convention yields results for the activity coefficient given in Fig.(\ref{fig:13})

\begin{figure}[htbp]
\begin{center}
\includegraphics[ width=11cm]{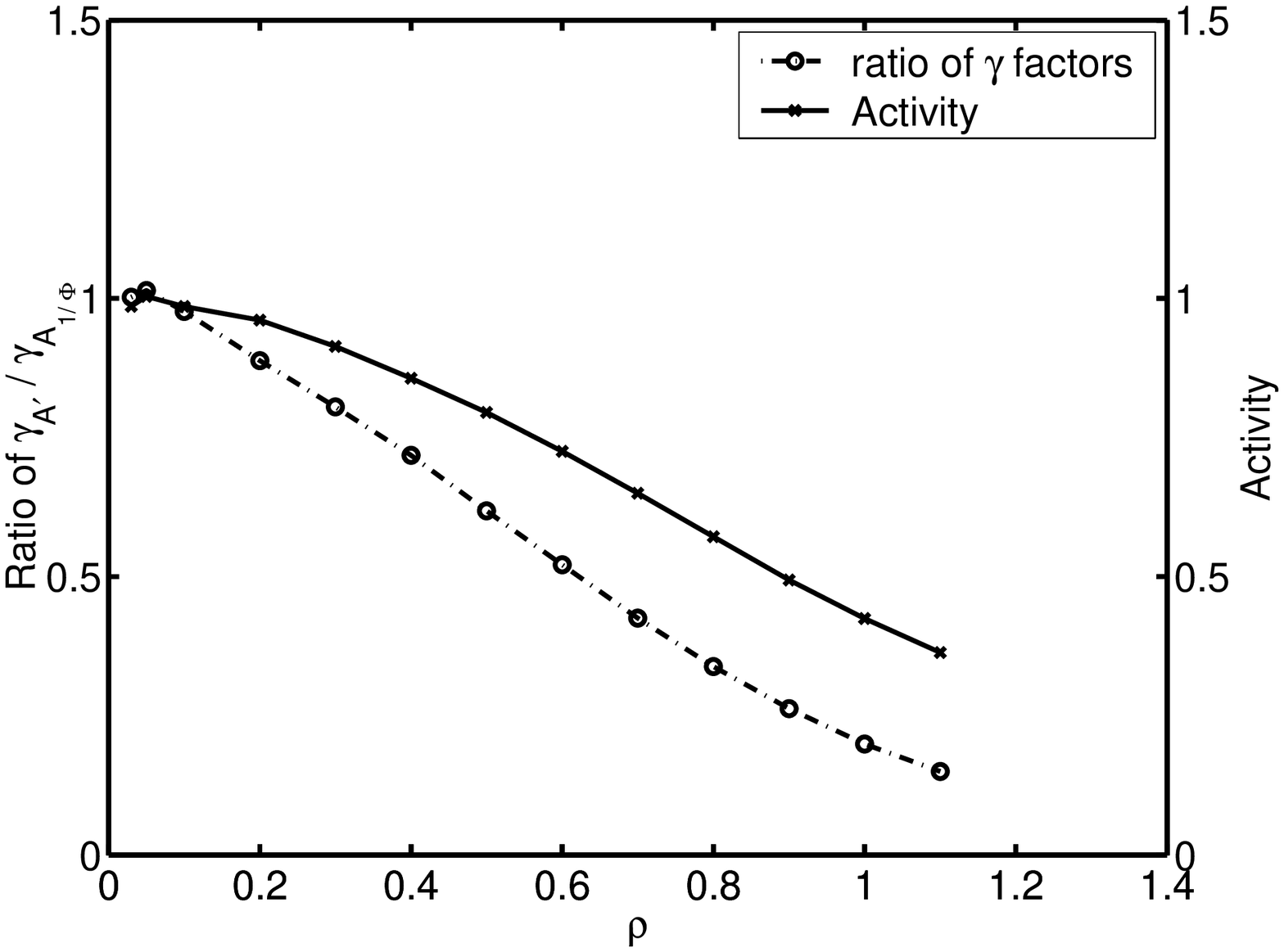} 
\end{center}
\caption{Availability coefficient ratios to illustrate the expected constancy of the availability coefficients of the excited species.}
\label{fig:29} 
\end{figure}

It is of interest to determine the  activity
coefficients by direct calculations because  the Kwong-Redlich
and associated equations \cite[p.29]{hill2},\cite{red1}
used to determine activity coefficients for multicomponent mixtures
are not easily extended to  chemical reactions: part of the difficulty is that 
the component concentrations in their thermodynamical mixtures are
free to vary independently, and may be fixed at any arbitrary value:
this cannot obtain for reactive systems because of the relational  dependence
of the components via the equilibrium constant, and partial derivatives
of thermodynamical quantities used to determine these coefficients
demand free variation or fixing of concentration terms that is not
realizable in a reactive system. Moreover there is the need to determine the variation of
 $c(\rho)$  over the entire
system density because of its theoretical significance linking $\Phi$ and $\Phi'$. The next section provides a broad methodology.  the exact methodology appears not to have been worked
out but verification of the above  would constitute another
method of determining activity coefficients from rate measurements.

\begin{figure}[htbp]
\begin{center}
\includegraphics[ width=11cm]{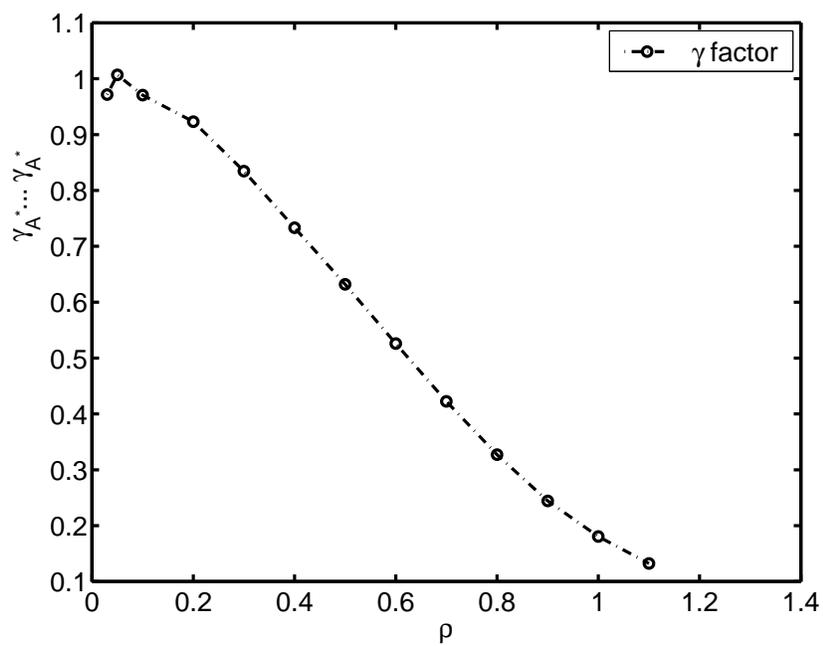} 
\end{center}
\caption{Excited state activity for dimer before bonding for forward rate  showing relatively  lower change compared to the bulk activity coefficients (which would be the square of the values given in Fig.~(\ref{fig:12})). }
\label{fig:32} 
\end{figure}
\subsection{Models based on  simulation results}\label{sub:3.5}
The elementary rate constants have the form \cite{hang1}
\begin{equation}\label{nw1}
k_i=A(T,\Omega) \exp\frac{-E}{kT}	
\end{equation}
where $E$ is termed the activation energy and $\Omega$ are variables intrinsic to the reaction, such as impact and structure parameters. The object here is to relate the $\gamma'$ reactivity coefficients to the $\gamma$ availability (activity) coefficient  based on the first order single or two stage energy perturbation mechanism (Figs.~ (\ref{fig:d2}-\ref{fig:d1})respectively. Refinements include perturbing the pathway itself, and introducing a continuous potential field along the entire length of the reaction pathway.  The object here is to present a broad framework where calculations to any degree of accuracy might be attempted. For both these mechanisms,  $A^\ast_2$ and $A^\ast$ are the states of the reactants just prior to product formation. The first order perturbation lifts the degenaracy  of the $\ast$ levels relative to the vacuum state; in the vacuum state, no singularities are observed in the potentials. Singularities due to the perturbation arises because the product and reactant states are distinct and distinguishable, and need not have the same activity in general; the form factor $A(T,\Omega)$ is not altered to first order since they specify the type of reaction; only $E$ is altered due to the external potential that modifies the ground state energies of the particles relative to the vacuum level. The availability coefficient $\gamma_X$ is expressed as 
\begin{equation} \label{nw2}
\frac{\epsilon_X}{kT}=\ln \gamma_X
\end{equation}
 for any species $X$, and the excitation energy is written $\delta B$ for species $B$.For what follows, $\epsilon$ always refer to an energy term associated with the activity or availability coefficient.  In the single stage model, the upper $X^\ast$ species is perturbed by an absolute  energy amount given by $\gamma_{X^\ast}$ in (\ref{nw2}); for the double stage model,the upper level is perturbed by the relative energy $\Delta E^0_{\{F,B\},2}$, due to the singularity induced by the change of state from product to reactant or vice-versa. The lower state for both models are perturbed (lessened) by the same factor $\gamma_X$ since this is the ground state availability . It will be shown that the 2-stage relative model is the more accurate and logical, based on a comparison with estimates of the activity coefficient for the monomer or atom $A$. The backward activation energy $\epsilon^b_{act}$ for $A_2\rightarrow A^\ast_{2,out}\rightarrow Product$ for either the 1 or 2 stage model can be written (superscript o refers to the vacuum state)
 \begin{equation} \label{nw3}
 -\epsilon^b_{act}= -\epsilon^0_{b}+\epsilon_{A_2} -\delta A_{2,out}
 \end{equation}
  where we also write $\delta A_{2,out}=\delta A_{b}.\delta A_{b}$. These terms refer to the upper state perturbation of the $A^\ast_{2,out}$ species.  The forward model yields 
  \begin{equation} \label{nw4}
   -\epsilon^f_{act}=- \epsilon^0_{f}+ \epsilon_{A} + \epsilon_{A}-\delta A_{2,in}
\end{equation}
where $\delta A_{2,in}$ refers again to the upper energy state energy perturbations.   Using (\ref{nw2}) to convert to availability, where $\delta X=kT\ln \gamma_X$, (\ref{nw3},\ref{nw4}) yield
\begin{equation} \label{nw5a}
\gamma'_A\gamma'_A=\frac{\gamma_A\gamma_A}{\gamma_{A_2,in}}.
\end{equation}

On the other hand, the exact experimental condition in  Observation(\ref{obs2}) can be written $\Phi=\Phi'$ where the prime refers to the same expression in $\gamma'_X$ where 
\begin{equation} \label{nw5b}
\gamma'_{A_2} =\frac{\gamma_{A_2}}{\gamma_{A_b}\gamma_{A_b}}=\frac{\gamma_{A_2}}{\gamma_{A_{2,out}}}.
\end{equation}
Hence, (\ref{nw5a}-\ref{nw5b}) leads to the exact result 
\begin{equation} \label{nw6}
\frac{\gamma_{A_{2,in}}}{\gamma_{A_{2,out}}}=\frac{\gamma_{A_{2,in}}}{\gamma_{A_b}\gamma_{A_b}}=1
\end{equation}
 or $\delta A_b .\delta A_b=\delta A_{2,in}$. In view of the fact that the product/reactant potential interface is completely different from the reactant/product potential interface for this hysteresis dimer, the result of (\ref{nw6}) is indeed remarkable and may be stated as a kinetic principle:
 \begin{prin}\label{p1}
 The perturbed energy required to promote reactants to products at the reactant/product potential interface is of the same magnitude as for the reverse transition of products to reactants at the product/reactant potential interface, even if these interfaces are discontinuous, i.e. are spatially and  energetically distinct, for elementary reactions in equilibrium. 
 \end{prin}
 Fig.~(\ref{fig:1}) shows that for this study the interfaces are distinct;"'reversible"' pathways have coincident interfaces.
\subsubsection{Two stage backward reaction model}\label{sub:3.6.1}
In view of (\ref{nw6}) we can write the exact form 
\begin{eqnarray}
      \gamma'_A & = & \frac{\gamma_A}{\sqrt{\gamma_{A_2,\,out}}}    \\ \label{nw7a}
      \gamma'_{A_2} & = &  \frac{\gamma_{A_2}}{\gamma_{A_2,\,out}}   \label{nw7b}
 \end{eqnarray}
If a complete characterization of the $\gamma_{A_2}$ species were known, then $\gamma_{A_{2,\, out}}$ could be calculated from the theory presented in Sec.~(\ref{subsub:3.3.0}) especially expressions (\ref{eg7}). Some estimates or approximations may be made to determine the $\gamma$'s. The reaction pathway for this 2 stage backward step model may be written 
\begin{equation} \label{nw8}
Reactants (a)\rightarrow T.S. reactants (b)\rightarrow T.S. \,products (c) 
\end{equation}
When considered in isolation, (normal vacuum state analysis), there is continuity of the potentials between states (b) and (c) in (\ref{nw8}). On the other hand, if the environment is considered where species are either reactants or products, singularities in the energetics would develop about the same arbitrary volume of the $TS$ of the reaction coordinate; for instance at (b),  two atoms in proximity $A...A$ are separated by distance $\delta r_{A,A}$ and the mean activity  of this state is strongly  influenced  by the potential $V(\delta r_{A,A})$ dependent on the $\delta r_{A,A}$ distance, in addition to the surrounding non-participatory environmental molecules $V'(r_{others})$. On the other hand, at the same vicinity  (c) where the switch modifies the potentials, then the activity of the instantaneously formed molecule would be determined by $V'(r_{others})$, where $V(\delta r_{A,A})$ now represents the potential of internal coordinates not connected to the activity. An apparent discontinuity arises according to this first order treatment, which has no apparent analog in the traditional or other QM theories computed at vacuum densities. For the two stage reaction, two types of approximations for activity coefficients may be made for the forward (F) $2A\rightarrow A_2$  and backward (B) $A_2\rightarrow 2A $ kinetic pathways and the results may be checked with  the apparent activity coefficient values derived from the LJ single phase fluid.\newline
{\bfseries Case B:}
The first order perturbation energy $E(\gamma_{A_{2,\,out}})$ is 
\begin{equation} \label{nw9}
E(\gamma_{A_2,out}=2\epsilon_{A_b}-\epsilon^\ast_{A_2}
\end{equation}
where $A_b$ denotes the product atom just dissociated from the dimer at coordinate of species $A^\ast_2$. The following approximation will be derived;
\begin{equation} \label{nw10}
\delta A_{2,\, out}=\frac{\gamma_{A_b}\gamma_{A_b}}{\gamma_{A_2}}
\end{equation}
where $\gamma_{A_b}\approx \gamma_A$. Clearly, $\delta A_{2,\, out}$ serves as a "`retardant"' to the rate for positive $\epsilon$'s in the numerator of (\ref{nw10}).
Fig.~(\ref{fig:d1}) can be correlated with (\ref{nw8}) as follows: $E^0_{B,1}$ is the activation energy to the level $A_2^\ast$, which consists of  two degenerate states relative to the vacuum denoted $TS (b)$  with molecule $A_2$, and $TS (c)$ which refers to the two atoms $2A$ after dissociation; in real time the molecule disintegrates along the trajectory $(a) A_2\rightarrow (b) \rightarrow (c)$ according to (\ref{nw8}).  Unlike the atom, the dimer state may be characterized according to the inter particle $A-A$ distance $r_{A-A}$, (as well as the magnitude of the external potential). The mean external forces acting on this dimer would lead to a potential which is to a first 
approximation relatively invariant, even if the internal potential and internuclear distance varies; thus we assign $\epsilon (A_2(a))\approx \epsilon (A_2(TS,b)$ . The potential energy is utilized to overcome the activation barrier; the boundary  $\partial C$ imposes a fixed density on the system which elevates the mean potential of the reactants relative to the vacuum ground state. During the transition to the state $TS(b)$, external forces with a mean potential $\epsilon(A_2(TS)$ would operate on the dimer; $r_b=1.20$ for this reaction, where for a b.c.c. approximation for a LJ fluid at $\rho=0.70$, the approximate nearest neighbor distance is $1.22$, implying that the environment of $A(TS)$ is very close to that of the bulk  or generic $A$ atom, i.e. a typical atom with activity coefficient $\gamma_A$, so that the transition state $A_2$ availability coefficient for the atom  $\gamma_{A,TS}$ can be equated with the bulk,  $\gamma_{A,TS}\approx \gamma_A$ and the relative energy to be overcome from this $TS$ state is $\Delta E^0_{B,2}$ where $\Delta E^0_{B,2}\approx (2\epsilon_A(T.S.)-\epsilon_{A_2}(TS) )$ leading to 
\begin{equation} \label{nw11}
\exp-\frac{\Delta E^0_{B,2}}{kT} \approx \left(\frac{\gamma_{A_2}}{\gamma_A\gamma_A}\right).
\end{equation}
Because $\exp-\Delta E_{B,1}= \exp -\Delta E^0_{B,1}\gamma_{A_2}$, we have the forms for the backward rate given as 
\begin{eqnarray}
   k_B    & \approx & k^0_B \nonumber \gamma_{A_2}\frac{\gamma_{A_2}}{\gamma_{A,TS}\gamma_{A,TS}}\\ 
       &\approx   &  k^0_B \gamma'_{A_2}\nonumber\\ 
       &\approx&  k^0_B \gamma_{A_2}\Phi=k^0_B \frac{\gamma_{A_2}}{(1/\Phi)} .\label{nw12}
 \end{eqnarray}

But (\ref{nw11}-\ref{nw12}) implies 
\[ \gamma'_{A_2}=\frac{\gamma_{A_2}}{\gamma_{A_2},in}\Rightarrow \gamma_{A_2,in}=\gamma_{A_2,out}\approx \frac{1}{\Phi} \] leading to the useful approximation
\begin{equation} \label{nw13}
\gamma_{A_2}\approx \frac{\gamma'_{A_2} }{\Phi}
\end{equation}
From (\ref{nw7b}), $\gamma_{A_2}=\gamma'_{A_2}\gamma_{A_2,out}$ so that (\ref{nw10}) follows with $\gamma_{A_b}\approx \gamma_A$.
Since $\gamma_{A_2,in}=\gamma_{A_2,out}$, and $\gamma'_A\gamma'_A=\frac{\gamma_A\gamma_A}{\gamma_{A_2,in}}$, we get
\begin{equation} \label{nw14}
\gamma_A \approx \frac{\gamma'_A}{\sqrt{\Phi}}.
\end{equation}
Since the r.h.s. of (\ref{nw13}-\ref{nw14}) are available, we plot $\gamma_A$ and $\gamma_{A_2}$ in Fig.~(\ref{fig:12}). We expect $\gamma_A$ to be close or at least follow the trends of those derived from the  Free Energy estimates given in Figures(\ref{fig:14}-\ref{fig:16}), especially Fig.~(\ref{fig:14}). The other two figures were plotted to show that the coefficient estimates  are all greater than unity ($>1$). As expected from the theory provided, the activity coefficients from the $A_{res}$ function show semi-quantitative agreement with the simulation results. The other $A_{res}$ functions which eliminate the dimer contribution shows markedly lower values; we would expect a slight lowering of value due to dimer interference, leading to the conclusion that simulation results with the above two models are plausible, and that $A_{res}$ is the more appropriate function to use for estimation. Since the adjacent atom in the dimer cannot contribute to the potential energy of the dimer contributing to its activity, one would expect its coefficient to be lower. The (B) reaction above has its counterpart in the forward (F) reaction, but because the transition state atomic activity coefficient is severely affected by its own force acting on the other target atom, it would not be reasonable to assume that the activity coefficient at that $TS$ state due to the particle potentials can be equated with the bulk activity, especially when the internuclear distance at $r_f=0.85$ corresponding to an energy of over $17.0$ LJ units! However it would be instructive to follow through the consequences of this approach so that comparisons with the activity coefficients derived from the literature may be made. \newline
{\bfseries F.Process} Using the same arguments as for the B process, Fig.~(\ref{fig:d1})
(F process) gives the net elevation of the vacuum  levels about $\Delta E^0_{F,2}$ such that 
\begin{equation} \label{nw15}
\exp-\frac{\Delta E^0_{F,2}}{kT} \approx \left(\frac{\gamma^\ast_{A_2}}{\gamma^\ast_A\gamma^\ast_A}\right).
\end{equation}
The ground state relative to the vacuum is elevated by $\epsilon_A$ per particle, so that 
\begin{equation} \label{nw16}
E_{F,1}=E^0_{F,1}-2\epsilon_A
\end{equation}
where as usual the $\epsilon$'s to the potential energy associated with the availability coefficients.
The total activation energy $\Delta E_{F,tot}$ is then given by
\begin{equation}\label{nw17}
\Delta E_{F,tot}=E_{F,1}+ \Delta E^0_{F,2}	
\end{equation}
By the definition of the reactivity coefficients,
\begin{equation} \label{nw18}
\gamma'_A\gamma'_A=\exp\left[\frac{2\epsilon_A-\Delta E^0_{F,2}}{kT}\right]
\end{equation}
 and (\ref{nw18}) leads to 
 \begin{equation} \label{nw19}
\gamma '_A \gamma '_A  = {\raise0.7ex\hbox{${\gamma _A \gamma _A }$} \!\mathord{\left/
 {\vphantom {{\gamma _A \gamma _A } {\left( {\frac{{\gamma _{A_2 }^* }}
{{\gamma _A^* \gamma _A^* }}} \right)}}}\right.\kern-\nulldelimiterspace}
\!\lower0.7ex\hbox{${\left( {\frac{{\gamma _{A_2 }^* }}
{{\gamma _A^* \gamma _A^* }}} \right)}$}}
\end{equation}
where the asterisked states are at the $TS$. Despite the arguments above, if we were to make the approximation $\gamma^\ast_A \approx \gamma_A$ (where in general , only a detailed perturbation theory would yield an appropriate value of   $\gamma^\ast_A)$ then
$\gamma^\ast_{A_2}\approx \gamma_{A_2}$ and $\gamma^\ast\approx \gamma_{A}$ yields 
\begin{equation} \label{nw20}
\frac{\gamma^\ast_{A_2}}{\gamma^\ast_A\gamma^\ast_A}\approx \Phi
\end{equation}
Then (\ref{nw19}) implies 
\begin{equation} \label{nw21}
\gamma'_A \gamma'_A=\frac{\gamma_A \gamma_A}{\Phi}
\end{equation}
yielding
\begin{eqnarray}
     \gamma_A  & \approx  & \gamma'_A\sqrt{\Phi}\label{nw22a}\\
         \gamma_{A_2}   & \approx   &\Phi \gamma'_{A_2}\label{nw22b}
 \end{eqnarray}
since here $\Phi=\Phi^\ast$. The above so-called $\Phi$ convention plots for the activity coefficients is given in Fig.~(\ref{fig:13}) where $\gamma_A,\gamma_{A_2}< 1$, which contradicts the results of  the free energy estimates given in Figs.~(\ref{fig:14}-\ref{fig:16}), especially Fig.~(\ref{fig:14}); this was to be expected from the assumptions made for state $A^\ast$. Since $\delta A_{2,out}$ can be determined from the more reliable (B) process, we might be able to derive an estimate for $\gamma^\ast_A\gamma^\ast_A$ for the $(A...A)$  \,$TS$ state of two atoms about to dimerize where the (B) process assignment is used for the determination. We would expect $\gamma^\ast_A\gamma^\ast_A$ to change {\itshape relatively} much more slowly compared to the $\gamma_A$ and $\gamma_{A_2}$ variations over the system $\rho$ if the activity coefficient is strongly influenced by the isolated large repulsive potential of the two atoms at the $TS$ region. Evidence in this direction would serve as a prototype for describing ionic reactions along the classical ideas of Br{\"o}nsted and Bjerrum.

\subsubsection{Single-stage considerations} 
There is no reference state at the $TS$ to compute energy differences. For instance, for the backward reaction
\begin{equation} \label{nw23}
\gamma'_{A_2}=\frac{\gamma_{A_2}}{\gamma^\ast_{A}}{\gamma^\ast_{A}}\approx \Phi
\end{equation}
if $\gamma^\ast_A\approx \gamma_A$ as before for (B) (2 stage process). Then from $\Phi=\Phi'$ we get $\gamma'_A\approx 1$ which is contrary to the simulation results.
Similarly, for the (F) reaction, we have $\gamma'_A \gamma'_A =\frac{\gamma_A \gamma_A}{\gamma^\ast_{A_2}}$ and with the assignment $\gamma^\ast_{A_2}\approx \gamma_{A_2}$, we get 
$\gamma'_A =\frac{1}{\sqrt{\Phi}}$ and from (\ref{nw7b}) there results $\gamma'_{A_2}\approx \frac{\gamma_{A_2}}{\gamma_{A_2}}=1$ which is also not observed. Hence the absolute single-stage process are not expected although they seem to conform to the Br{\"o}nsted-Bjerrum (B-B) form (where the $z$'s are the associated charges)
\[A^{z_a} +B^{z_b} \rightarrow (AB)^{z_a +z_b}\rightarrow Products   \]
with $k=k^0\Delta$ where $\Delta =\frac{\gamma_{A^{z_a}} \gamma_{B^{z_b}}}{\gamma^{\dag}_{AB^{(z_{a}+z_b)}}}$ with $\dag$ referring to the upper level in the energy diagrams of Figs.~(\ref{fig:d2}-\ref{fig:d1})or transition state of conventional theory. These observations suggest that the B-B form could well be described as an approximation of another mechanism which does not have pre-equilibria transition states; an example being the two stage process below which subsumes the B-B equations. 
\newline{\bfseries{Ionic Reactions:}} The classical kinetic theory of salt effects applied to recent investigations by Sanchez et al. \cite{san1,san2} gives the B-B equation for the rate constant $k$ as 
\begin{equation} \label{nw24}
k=k_0 \frac{\gamma_A\gamma_B...}{\gamma_\dag}
\end{equation}
where $k_0$ is the reference rate constant for the solvent, $\gamma_A,\gamma_B ...$ are the reactant activity coefficients of reactants $A,B,...$ and $\gamma_\dag$ is that of the transition state where if the charge of the reactants are $z_a, z_b,...$ then the charge of the $TS$ denoted  by $\dag$ is $z_\dag=z_a+z_b+....$. The activity coefficient $\gamma_J$(Debye-Huckel limiting law approximation) for any species $J$ conforms to 
\begin{equation} \label{nw25}
\log \gamma_J=-Az^2_JI^{1/2}=-Qz^2_J
\end{equation} 
where $Q=AI^{1/2}$ is a positive number, leading from $(\ref{nw24})$ to the rate form
\begin{equation} \label{nw26}
\log k_2=\log k^0_2 +2Az_Az_B I^{1/2}
\end{equation}
where a negative salt effect is expected for anion/cation bimolecular reactions, such as the recently studied reaction (with pz=pyrazine)
\begin{equation} \label{nw27}
\underbrace {\mbox{[Fe(CN)$_6$]}^{3-}}_{A}+\underbrace {\mbox{[Ru(NH$_3$)$_5$pz]}^{2+}}_{B}\,\, \stackrel{\stackrel{\scriptstyle k_r}{\rightleftharpoons}}{\scriptstyle k_f}\,\, \underbrace {\mbox{[Fe(CN)$_6$]}^{4-}}_{C} + \underbrace {\mbox{[Ru(NH$_3$)$_5$pz]}^{3+}}_{D}
\end{equation}
where the forward reaction with rate constant $k_f$ shows a contradictory positive salt effect; other violations have been reported \cite[see ref. 4 of this citation]{san2}. The interpretation of these anomalies has been made using the theory of Marcus and Hush \cite{san2}.  These workers introduced composite reactions, leading to a pseudo-elementary process such as 
\begin{eqnarray}
  A+B     & \stackrel{\stackrel{\scriptstyle k_1}{\rightleftharpoons}}{\scriptstyle k_{-1}} &\,\, PC \nonumber \\
    PC   & \rightleftharpoons  & X^\dag\rightleftharpoons\,\,SC \nonumber\\
   SC \,\, &\rightarrow& \,\,C+D \label{nw28}
 \end{eqnarray} 
for the overall reaction $A+B\rightarrow C+D$. Violations of the B-B formula (\ref{nw24}) have been attributed to differences in activity coefficient properties of the precursor complex $PC$ and activated state $X^\dag$; $SC$ is a postulated "`successor complex"', where the Markus-Hush ideas are also incorporated \cite[p. 15089]{san2}. It would be of interest therefore to frame  theories and proposals  for elementary reactions that might also explain some of the above results; this is attempted below.
\subsubsection{(i) Proposal of mechanism to explain positive deviation for forward reaction of eq.(\ref{nw27})}\label{sub:3.6.3}

Rewriting (\ref{nw27})with reactants $C$ and $D$, products $A$ and $B$  with the charges, we have an elementary reaction 
\begin{equation} \label{nw29}
C^{4-}+ D^{3+}\stackrel{k_f}{\rightarrow}A^{3-} + B^{2+}
\end{equation}
The postulated reaction sequence route of an ideal model is  
\begin{equation} \label{nw30}
C^{4-}...D^{3+}\rightarrow T.S.^{\mbox{l state}}\rightarrow \left\{A^{3-}+B^{2+}\right\}^{\mbox {u state} }\,\,\mbox{separated products P}
\end{equation}
In (\ref{nw30}), $C...D$ represents a {\itshape pre-associated} complex which retains the spectroscopic details (e.g. in UV-IR range used for measuring concentrations) but which are close enough to ideally form an effective single charge of magnitude $-1$. The $T.S.$ is the lower (l) degenerate state relative to the vacuum, and P the upper level $u$. In the presence of the solution dielectric, the first order perturbation according to the two stage model give a total pre-exponential total $\gamma_{tot}$ as 
\begin{equation} \label{nw30.1}
\gamma_{tot}=\gamma_{C...D}\left/ \left(  {\frac{\gamma_A \gamma_B\,\,(\mbox{u state})}{\gamma_{T.S.}(\mbox{l state})}} \right)    \right.
\end{equation}
for the overall rate $k_2=k^0_2\gamma_{tot}$. Taking (\ref{nw25})logarithms to base 10,  results in
\begin{eqnarray}
   \log  \gamma_{tot}   &  = & \log \gamma_{C...D} +\log \gamma_{T.S.}-\log \gamma_{A}-\log \gamma_{B} \label{nw31}\\
       & =  & -1Q  -1Q + 9Q + 4Q   \nonumber \\
       &\approx&+11Q \label{nw32}
 \end{eqnarray}

and so 
\begin{equation} \label{nw33}
\log k_2=\log k^0_2 +11Q
\end{equation}
which leads a positive deviation in contradiction to the standard B-B equation for cation/anion reactants. The above model had no intermediate forms; all charges were discrete etc.. AS a further refinement, one can write down intermediate forms of association with charge parameter $-\lambda$ and the $l$. state with charge $-\tau$ leading to 
\begin{equation}\label{nw50}
	\log k_2=\log k^0_2 +Q(13-\lambda^2 -\tau^2)
\end{equation}
  It might be possible to write down intermediate forms of association along the lines of the above where the positive deviation is not a discrete multiple of $Q$ above.
\subsubsection{(ii) Other mechanisms not in B-B form}\label{sub:3.6.4}
Large charged molecular fragments $A^{z_a}$ and $B^{z_b}$ might not be large enough to be considered separate and discrete even at the TS region so that for the overall elementary reaction
\begin{equation} \label{nw34}
A^{z_a} + B^{z_b} \rightarrow Products
\end{equation}
with  pathway $A+B \rightarrow T.S.\rightarrow P$, the two stage proposal yields the pre-exponential term $\gamma_{tot}$ to $k_2$ where 
\begin{equation} \label{nw34b}
\gamma_{tot}=\gamma_A\gamma_B \left/ \left(  \frac{\gamma_P}{\gamma^{T.S.}} \right)    \right..
\end{equation}
If there is no further fragmentation, the charge on $P$  would equal that of the $T.S$  located at two centers, leading to $\gamma_{tot}=\gamma_A\gamma_B$ with 
\begin{equation} \label{nw35}
k_2=k^0_2\gamma_A\gamma_B.
\end{equation}
On the other hand, $A$ and $B$ if distinct would have the $T.S.$ activity coefficient $\gamma_{T.S.}\approx \gamma_A.\gamma_B$ and if $P$ fragments, then the $\gamma_{tot}$ multiplicative factor becomes 
\begin{equation} \label{nw36}
\gamma_{tot}=\left( \gamma_A\gamma_B\right )^2\left/ \gamma_P\right .
\end{equation}
where the $\gamma_P$ is a product of the activity coefficients of the fragmented portions of the product states. 
\subsubsection{(iii) General harmonization of the B-B formulation with the present theory for free (non-associated) ions}\label{sub:3.6.5}
The ionic reaction with free ions 
\[ A^{z_a} + B^{z_b} \rightarrow Products \,\, (P)\]
has the following reaction pathway for the two stage model  
\begin{equation} \label{nw38}
A^{z_a} + B^{z_b}\rightarrow A^{z_a}...B^{z_b}\,(T.S.(a))\stackrel{\Delta \mbox{E=0,vacuum}}{\rightarrow} (A+B)^{z_a+z_b}\,(T.S. (b))\rightarrow P.
\end{equation}
$\Delta E_{act}$ is defined to be the activation energy parameter from the free ionic state to the transition state $a$ under vacuum conditions where there is bond formation  and the smallest distance $\delta r_{A,B}$ between two charge centers that are considered separate, and T.S. $b$ is the smallest distance when they are not considered separate and when a singularity of the external potential is applied as in the two stage model. Then the first order energy perturbation at the TS (between ($a$) and ($b$)) $\Delta E$ is 
\begin{equation} \label{nw39}
\Delta E=kT \ln \left( \frac{\gamma_{(A+B)^{z_a+z_b}}} {(\gamma_{A^{z_a}}...\gamma_{B^{z_b}})}\right) 
\end{equation}
and so the rate constant $k_2$ becomes outside of the vacuum state 

\begin{equation} \label{nw40}
k_2=k_2^0 \frac{\gamma_A\gamma_B}{\gamma_{(A+B)^{z_a+z_b}}}.(\gamma_{A^{z_a}}...\gamma_{B^{z_b}})
\end{equation}

where $(\gamma_{A^{z_a}}...\gamma_{B^{z_b}})$ is the $TS$ species which has distinct charges relative to the dielectric medium and would appear as a product term of activity coefficients at the $TS$ state, i.e. $(\gamma_{A^{z_a}}...\gamma_{B^{z_b}}) =\gamma^{\ast(T.S)}_A.\gamma^{\ast (T.S)}_B$. The energy of interaction of the $A^{z_a}$ and $B^{z_b}$ would constitute a major portion of the activity coefficient of  the pair, and would be expected to be \emph {relatively} constant over a large range of system $\rho$ compared to the bulk activity coefficients of the other species and reactants. Thus,under these conditions, we have 
\begin{equation} \label{nw41}
k_2\approx k_2^{0'} \frac{\gamma_A\gamma_B}{\gamma_{(A+B)^{z_a+z_b}}}
\end{equation}
with $k_2^{0\prime}=k_2^{0}(\gamma_{A^{z_a}}...\gamma_{B^{z_b}})$, which is the B-B equation for ionic reactions. Clearly, some demonstration of the relative constancy of $\gamma_{A^{z_a}}...\gamma_{B^{z_b}}$ would be re-assuring. Obviously the potentials for the present dimer model and that of ionic reactions are different, but we can show a relative constancy for the analogous  $TS$ atomic pair, and so can expect a similar relative constancy to obtain for ionic reactions. We return to  the previous system of eq.(\ref{nw19}) where $\gamma_{A^{z_a}}...\gamma_{B^{z_b}}$ is equivalent to $\gamma^\ast_A\gamma^\ast_A=(\gamma^\ast_A...\gamma^\ast_A)$ for $A=B$.
Then 
\begin{equation} \label{nw42}
\gamma'_A \gamma'_A = \frac{\gamma^2_A(\gamma^\ast_A...\gamma^\ast_A)}{\gamma^\ast_{A_2}}  
\end{equation}
A plot of this function is given in Fig.~(\ref{fig:32}). The analog of $\gamma^\ast_{A_2}$ is $\gamma_{(A+B)^{z_a+z_b}}$. In (\ref{nw41}), for $A=B$, $\gamma_A \gamma_B \equiv \gamma_A^2$ would vary very dramatically, as is obvious from Fig.~(\ref{fig:12}). From $\rho=0$ to $\rho=1$, $\frac{\gamma^2_A}{\gamma_{A_2}}$  varies by about $16$, whereas $(\gamma^\ast_A...\gamma^\ast_A)$ varies by  $0.2$. Thus, it is conceivable that the B-B expression may be derivable from a strictly elementary reaction, based on the above estimates.Fig.~(\ref{fig:29}) illustrates two plots, where the "Activity"' variable corresponds to  using (\ref{nw42}) to calculate $\gamma_A^\ast$ with the   assumption $\gamma_{A_2}^\ast=\gamma_{A_2}$ and the "`Ratio"' variable is simply the reactivity and activity ratio $\frac{\gamma'_A}{\gamma_A}$ using the inverse $\Phi$ convention to determine $\gamma_A$.
\section{Conclusion}\label{sec:4}
The current form of the elementary reaction rate constant is still incomplete despite the depth and complexity of analysis valid for relatively low density systems due to the utilization of traditional definitions and conventions. A more complete description would involve the "`reactivity"' coefficients and these coefficients are intimately related to the activity coefficients of the species involved in the reaction. Under stipulated conditions, such as obtain for the two stage model, explicit first order expressions may be derived connecting the reactivity and activity coefficients. A whole range of product and reactant states exists, and under an enumeration scheme incorporating the Gibbs equilibrium condition, one particular state may be determined from another through a series of equations which also feature the bulk activity coefficients.
The theory of ionic reactions as developed by Br\"onsted and Bjerrum  may be subsumed by the simple first order models provided here using a strictly elementary reactive process; in particular these models can also explain apparent violations of the standard Br\"onsted-Bjerrum theory involving composite reactions. The  framework given here can be extended to  other more complex particle reactions which are moderated by strong external fields, such as what might obtain in the  plasma state. 

\textbf{Acknowledgement} C.G.J would like to thank Jayendran C. Rasaiah (Chemistry Dept.,University  of Maine (Orono) U.S.A.) for his generous hospitality there over a two month period during which time the bulk of  this work was completed and the remnant of I.R.P.A. (Malaysian)  grant no. 09-02-03-1031 for financial assistance. 

\bibliographystyle{unsrt}
\bibliography{react3}

\end{document}